\documentclass[12pt,letterpaper]{article}
\pdfoutput=1
\usepackage{jheppub}
\usepackage{amsfonts,amsthm}
\usepackage[english]{babel}
\usepackage[utf8]{inputenc}
\usepackage{indentfirst}
\usepackage{slashed}
\usepackage{mathrsfs}
\usepackage{amssymb}
\usepackage{color}
\hypersetup{unicode}
\usepackage{natbib}
\usepackage{mathrsfs}
\usepackage{array}

\newcommand{\ma}[1]{\mbox{$\mathcal{#1}$}}

\newcommand{\mrm}[1]{\mbox{$\mathrm{#1}$}}

\usepackage{verbatim}

\newcommand{\ie}{\emph{i.e.}}

\DeclareMathOperator{\tr}{tr}

\newcommand{\ZZ}{\mathbb{Z}}

\newcommand{\RR}{\mathbb{R}}

\newcommand{\mr}[1]{\mathrm{#1}}

\newcommand{\U}[1]{\mathrm{U}(#1)}
\newcommand{\SU}[1]{\mathrm{SU}(#1)}
\newcommand{\SO}[1]{\mathrm{SO}(#1)}
\newcommand{\USp}[1]{\mathrm{USp}(#1)}
\newcommand{\Spin}[1]{\mathrm{Spin}(#1)}

\newcommand{\AdS}{\mathrm{AdS}}
\newcommand{\vol}{\mathrm{vol}}

\mathchardef\mhyphen="2D

\title{Defects in 4d SCFTs from supergravity and holographic renormalization}

\author[a]{Federico Faedo}
\author[b]{Nicol\`o Petri}
\author[a]{Alessia Segati}

\affiliation[a]{International Centre for Theoretical Physics Asia–Pacific,
University of Chinese Academy of Sciences, 100190 Beijing, China.}
\affiliation[b]{INFN sezione di Torino, via Pietro Giuria 1, 10125 Torino, Italy.}

\emailAdd{federico.faedo@ucas.ac.cn}
\emailAdd{nicolo.petri@to.infn.it}
\emailAdd{alessia.segati@ucas.ac.cn}

\abstract{Type IIB string theory admits Janus solutions connecting AdS$_5$ vacua. These geometries correspond to interface conformal field theories and are naturally described in five-dimensional supergravity as AdS$_4$ foliations over an interval. In this work, we introduce a novel type of Janus solution in AdS$_5$, featuring AdS$_2 \times S^2$ foliations instead of AdS$_4$. We interpret these solutions as the supergravity duals of line defects in four-dimensional SCFTs and we discuss their D-brane engineering.
Our primary tool is Romans supergravity in five dimensions, which provides a simple Type IIB truncation over $S^5$. Within this framework, under specific conditions, we construct regular solutions interpolating between AdS$_5$ and AdS$_2 \times \mathbb{R}^3$. These geometries represent one-parameter deformations of AdS$_5$ vacua and are characterized by a running profile for two-form gauge potentials and a scalar field in the supergravity bulk. Finally, using holographic renormalization, we explicitly compute the on-shell action of the defect and we discuss the one-point functions in the holographic field theory.}

\begin{document}
\maketitle
\flushbottom

\section{Introduction}
A key insight from string theory is that quantum gravitational systems lack free parameters. This idea appears in various forms in the literature, such as the notion that every coupling constant corresponds to the asymptotic value of a dynamical field or that every symmetry must be gauged \cite{Banks:2010zn}. In this context, 2D gravity stands out as an exception (for a very short list of relevant results on this topic, see \cite{Polchinski:1989fn,Maldacena:1998uz,Strominger:1998yg,Hartman:2008dq,Maldacena:2016hyu, Maldacena:2016upp,Saad:2019lba}). In this regard, it seems reasonable to assume that Swampland criteria apply only to quantum gravities in dimensions $D\geq3$. Even in $D=3$, gravity exhibits notable peculiarities, as it lacks dynamical degrees of freedom. Consequently, this restriction is often further refined to apply only when $D>3$ \cite{McNamara:2020uza}.

Despite the extensive literature on these topics, many aspects of quantum gravity in two and three dimensions remain elusive and pose significant challenges. A key issue lies in understanding the UV origin of two-dimensional gravitational configurations and their consistent embedding within string theory. One of the greatest theoretical obstacles in 2D gravity is providing a precise formulation of the holographic principle or, equivalently, understanding the deep meaning of the sum over different topologies (see, for instance, \cite{Harlow:2018tqv,Marolf:2020xie,Eberhardt:2021jvj,Gesteau:2024gzf}).

This paper follows this general research trajectory, focusing on supersymmetric AdS$_2$ and its embedding in string theory. Using the standard AdS/CFT framework, we explore the idea that AdS$_2$ can be associated holographically to line defects within higher-dimensional superconformal field theories. Specifically, we will focus on four-dimensional SCFTs, as they are among the most studied and well-understood within the framework of holography and string theory. The notion of line defect in 4D SCFTs can be precisely developed in supergravity, starting from the crucial geometric property that the AdS$_5$ space can be locally written as a foliation of a AdS$_2\times S^2$ space. This research trajectory on defects in holography has proven fruitful, with many results in recent years focused on identifying low-dimensional AdS solutions dual to conformal defects within SCFTs in various dimensions. For a non-exhaustive list of references on AdS$_2$ and AdS$_3$ as holographic defects, see \cite{DHoker:2007mci,Chiodaroli:2009yw,Chiodaroli:2011fn,Bobev:2013yra,Dibitetto:2017klx,Gutperle:2017nwo,Dibitetto:2018iar,Dibitetto:2018gtk,Gutperle:2018fea,Chen:2019qib,Gutperle:2019dqf,Chen:2020mtv,Faedo:2020nol,Faedo:2020lyw,Chen:2020efh,Lozano:2021fkk,Lozano:2022ouq,Anabalon:2022fti,Lozano:2022swp,Lozano:2022vsv,Lozano:2024idt,Conti:2024qgx,Gutperle:2024yiz}.
Precisely, in this work we focus on the supergravity duals of line defects within four-dimensional SCFTs with $\mathcal{N}=2$ and $\mathcal{N}=4$ supersymmetry. For a very short list of references on defects in four-dimensional conformal field theories and holography, see \cite{Erdmenger:2002ex,Constable:2002xt,Kapustin:2005py,Gaiotto:2008sa,Gaiotto:2008sd,deLeeuw:2015hxa,Billo:2016cpy,Bianchi:2018zpb,Bianchi:2019sxz,Arav:2024exg}.

Defects in quantum field theory are among the most actively studied topics. Broadly, a defect in a field theory can be described as a deformation with a coupling constant locally defined in spacetime. In conformal field theories, a particularly interesting scenario occurs when the deformation preserves a lower-dimensional conformal symmetry, restricted to the spacetime directions where the defect extends. In holography and string theory, these ideas were initially explored in earlier works \cite{Karch:2000gx, DeWolfe:2001pq, Aharony:2003qf,Bachas:2001vj}. From this perspective, the defect describes the boundary conditions of the intersection between a stack of D-branes associated to a given SCFT and additional {\itshape defect branes}, which partially break the conformal isometries of the dual gravity vacuum. The lower-dimensional AdS geometry (AdS$_2$ in our case) provides the geometric picture of the locus of the intersection with defect branes.

Supergravity solutions that describe the fully backreacted geometries dual to conformal defects are the so-called Janus solutions \cite{Bak:2003jk,Clark:2004sb,Clark:2005te,DHoker:2006qeo,DHoker:2006vfr,DHoker:2007zhm,DHoker:2009lky,Aharony:2011yc,Bobev:2020fon,Ghodsi:2023pej,Conti:2024rwd,Arav:2024wyg}. These are domain wall geometries characterized by a foliation with a lower-dimensional AdS space, AdS$_2$ in this case, and a higher-dimensional AdS in the asymptotic region. Importantly, the fluxes associated with defect branes do not vanish asymptotically, reproducing only a locally AdS geometry. Another key feature of these solutions is that they usually form a one-parameter family, representing a deformation of the vacuum geometry. When this parameter is set to zero, the global vacuum is recovered, and all fluxes from the defect branes vanish. This parameter is ultimately related to the charges of the defect branes and, as we will see, plays a crucial role in determining the on-shell action of the defect.

In this work, we explicitly derive two new classes of AdS$_2\times S^2\times S^1\times S^3$ solutions fibered over two intervals in Type IIB string theory. We construct these backgrounds as Janus solutions in five dimensions and then uplift them to 10D employing the truncation formulas of \cite{Lu:1999bw}. The reduction used is a simple truncation over $S^5$, reproducing the so-called $\SU2\times \U1$ Romans supergravity in 5D. This is the {\itshape minimal} realization of $\ma N=4$ gauged supergravity in five dimensions, where ``minimal" indicates that only the supergravity multiplet is retained in the reduction. In this context, we obtain two inequivalent classes of AdS$_2\times S^2$ geometries fibered over an interval, both asymptotically locally AdS$_5$ and preserving 8 real supercharges (BPS/2 in 5D). Both families feature a supergravity bulk with a running dilaton and two two-form gauge fields. These excitations break the vacuum isometries and are related to interactions with defect branes.

The first class exhibits a spacetime singularity. In this case, we are able to relate the 10D uplift of our solutions to a class of near-horizon geometries of the type AdS$_2\times S^2\times S^1\times S^3$ fibered over two intervals, introduced in \cite{Lozano:2021fkk}. Specifically, we derive the explicit coordinate change leading to the 10D solution of \cite{Lozano:2021fkk}, that describes an intersection of D1-F1-D5-NS5 {\itshape defect} branes ending on a stack of D3s. This intersection breaks the isometries of AdS$_5\times S^5$ vacua to the singular backgrounds AdS$_2\times S^2\times S^1\times S^3$ fibered over two intervals. Equivalently, our 5D singular solutions can be also embedded within $\ma N=2$ AdS$_5\times S^5/\mathbb{Z}_k$ vacua, just by operating the topological substitution $S^3\rightarrow  S^3/\mathbb{Z}_k$ in the 10D uplift formulas. This situation corresponds to consider D3 branes over KK monopoles background.

The second class of solutions represents a completely new regular Janus configuration, characterized by a smooth limit where the 5D solution takes the form of an AdS$_2 \times \mathbb{R}^3$ geometry on one side and locally AdS$_5$ asymptotics on the other. These solutions form a one-parameter family of deformations of the AdS$_5$ vacuum, associated with a parameter $\lambda \in(-1,1)$. For this class, we provide the uplift in Type IIB, but we cannot determine a precise brane interpretation. However, we argue that the brane setup likely involves D3 branes with D1-F1-D5-NS5 defect branes. In this case, we suppose that the defect brane charges are {\itshape fully localized} in 10D spacetime, making it challenging to describe these solutions as near-horizon limits of a brane configuration.

Finally, for the regular Janus solution, we perform the precise computation of the on-shell action, using the prescriptions of holographic renormalization. This quantity encodes all the dynamical and geometric information about the defect degrees of freedom. Additionally, we compute the one-point function of the operator dual to the 5D scalar and we discuss the derivation of the one-point functions for the currents dual to the two-form gauge fields and the stress-energy tensor of the holographic field theory.

The paper is organized as follows: in sections \ref{AdS5branepicture} and \ref{sugrasetup}, we review the brane origin of AdS$_5$ vacua in string theory, 5D Romans supergravity, and the Type IIB consistent truncation. In section \ref{solutions}, we present the new 5D solutions, and in section \ref{IIBorigin}, we discuss their Type IIB uplifts. Section \ref{sec:holo-ren} contains the holographic computations of the on-shell action and one-point functions. In section \ref{conclusions}, we present our conclusions and future perspectives. Appendices \ref{app:gamma-matr} and \ref{app:counterterms} contain technical details regarding spinor technology for SUSY variations and the derivation of the counterterms needed for the computation of the on-shell action.

\section{The brane picture of AdS$_5$ vacua}\label{AdS5branepicture}

In this section we review some well-known properties of AdS$_5$ vacua in Type II supergravities and their brane interpretation. First, we consider the most simple setup of D4 and NS5 branes in Type IIA and we explicitly discuss how to extract $\ma N=2$ AdS$_5$ vacua in the near-horizon limit. We then consider T-dual configurations and discuss how AdS$_5\times S^5/\mathbb{Z}_k$ vacua arise from D3 branes on KK monopole background. The case with one KK monopole is the maximally supersymmetric vacuum AdS$_5\times S^5$.
We will mainly follow notations and approach of \cite{Lozano:2021fkk}, where the brane picture of AdS$_5$ solutions in Type IIA/B is extensively reviewed.

\subsection{AdS$_5$ from D4-NS5 intersections}

It is well-known that intersections of D4 and NS5 branes give rise to $\ma N=2$ gauge theories \cite{Witten:1997sc}. The main idea to engineer a 4d field theory from these branes is considering the D4s stretched between the NS5s. Schematically, we can depict these D4-NS5 intersections as in Table \ref{TypeIIAbranes}. More precisely, let us consider NS5 branes located within the space $\mathbb{R}_r^3$, parameterized by spherical coordinates $(r,\varphi^1,\varphi^2)$, and at some fixed values of the $\psi$ direction. Now take the D4s stretched between the NS5s along $\psi$. This implies that the D4s are finite in this direction, moreover D4 branes are located at a definite value within the plane parameterized by $(y, z)$. It then follows that the D4 worldvolume defines the color sector of a four-dimensional gauge theory. The gauge coupling is $g_4\sim \frac{\psi_{k+1}-\psi_k}{g_s}$, where the subscript $k$ identifies the position of NS5 branes. The theory is then conformal if the number of flavors\footnote{This setup can be enriched by including D6 flavor branes extended along $(\mathbb{R}^{1,3}, \mathbb{R}^3_r)$.} at each interval is equal to twice the number of colors \cite{Witten:1997sc}.
\begin{table}[http!]
\renewcommand{\arraystretch}{1}
\begin{center}
\scalebox{1}[1]{
\begin{tabular}{c | c cc  c|| c  c | c |c c c}
  & $t$ & $x^1$ & $x^2$ & $x^3$ & $y$ & $z$ & $\psi$ & $r$ & $\varphi^1$ & $\varphi^2$ \\
\hline \hline
$\mrm{D}4$ & $\times$ & $\times$ & $\times$ & $\times$ & $-$ & $-$ & $\times$ & $-$ & $-$ & $-$ \\
$\mrm{NS}5$ & $\times$ & $\times$ & $\times$ & $\times$ & $\times$ & $\times$ & $-$ & $-$ & $-$ & $-$ \\
\end{tabular}
}
\caption{The D4-NS5 system underlying 4d $\ma N=2$ SCFTs. The D4 are extended between NS5 branes along the direction $\psi$. The intersection is BPS$/4$ and admits warped AdS$_5$ vacua in the near-horizon.} \label{TypeIIAbranes}
\end{center}
\end{table}

Holographically, $\ma N=2$ superconformal theories are associated with infinite families of AdS$_5$ vacua in M-theory \cite{Gaiotto:2009gz}.
In this paper we will assume that the finite direction~$\psi$ is periodically identified. This condition implies that the $\ma N=2$ theories introduced above describe the $\mathbb{Z}_k$ orbifolds of the $\ma N=4$ SYM theory \cite{Witten:1997sc,Alishahiha:1999ds,Fayyazuddin:1999zu}. These theories have a very natural description in Type IIB in terms of the orbifold vacua AdS$_5\times S^5/\mathbb{Z}_k$.

Let's consider a supergravity description of the above Type IIA setup. We may start with the general metric in Type IIA associated to the D4-NS5 system
\begin{equation}
\label{NS5D4-TypeIIA}
d s_{10}^2 = H_{4}^{-1/2}  ds^2_{\mathbb{R}^{1,3}} + H_{4}^{1/2} \bigl(dy^2+dz^2\bigr)+H_{\mathrm{NS}5}H_{4}^{-1/2}\,d\psi^2+H_{\mathrm{NS}5}H_{4}^{1/2}\bigl(dr^2+r^2ds^2_{ S^2}\bigr)\,,
\end{equation}
where the Minkowski space $\mathbb{R}^{1,3}$ is parameterized by $x^0, \dots, x^3$.
The charge distributions of D4 and NS5 are encoded in the warp factors of the above metric. Specifically, D4 branes are taken fully-localized in their transverse space, \ie\ $H_{4}=H_{4}(y,z,r)$, while we assume the NS5 charge to be smeared along $\psi$, namely $H_{\mathrm{NS}5}=H_{\mathrm{NS}5}(r)$.
We may consider now the semi-localized solution defined as \cite{Youm:1999ti,Fayyazuddin:1999zu,Loewy:1999mn},
\begin{equation}\label{D4NS5solution}
 H_{4}=1+\frac{Q_{4}}{(y^2+z^2+4 Q_{\mathrm{NS}5}r)^2}\qquad\quad \text{and}\qquad\quad H_{\mathrm{NS}5}=\frac{Q_{\mathrm{NS}5}}{r}\,,
\end{equation}
where $Q_4$ and $Q_{\mathrm{NS}5}$ are integration constants related to D4 and NS5 charges.

This solution admits an AdS$_5$ vacuum geometry. In order to show this, the crucial observation is that a change of coordinates is needed. Let us define a new parametrization $(\mu,\alpha,\phi)$ as it follows \cite{Oz:1999qd}
\begin{equation}
  z= \mu \sin \alpha\,\sin\phi\,,\qquad\quad y= \mu \sin \alpha\,\cos\phi\,, \qquad\quad  r= 4^{-1}Q_{5}^{-1} \mu^2\cos^2\!\alpha\,,
\end{equation}
with $\alpha \in [0,\frac{\pi}{2}]$, $\mu>0$ and $\phi$ the angular polar coordinate in the $(y,z)$-plane. If we rewrite the solution \eqref{D4NS5solution} in this new parametrization and take the $\mu\rightarrow 0$ limit, we obtain the following geometry \cite{Oz:1999qd}
\begin{equation}
\label{D4NS5solutionAdS5}
d s_{10}^2 = Q_{4}^{1/2} ds^2_{\text{AdS}_5}+Q_{4}^{1/2}\left[d\alpha^2+4Q_{5}^2Q_{4}^{-1}c^{-2}d\psi^2 +s^2d\phi^2+4^{-1}c^2ds^2_{S^2}  \right]\,,
\end{equation}
where $c=\cos\alpha,\,\,s=\sin\alpha$ and  $ds^2_{\text{AdS}_5}=Q_{4}^{-1}\,\mu^2ds^2_{\mathbb{R}^{1,3}}+\frac{d\mu^2}{\mu^2}$.
This background is a particular realization of solutions \cite{Gaiotto:2009gz}. Specifically, the $R$-symmetry group of the dual theory can be read off from \eqref{D4NS5solutionAdS5} and is given by $\SU2\times \U1$ rotations acting on $S^2$ and $S^1_{\phi}$.

\subsection{The Type IIB picture}\label{ref:D3KKsystem}

We now perform a T-duality transformation along the circle $S^1_{\psi}$ in \eqref{D4NS5solutionAdS5}. From the supergravity side, it can be precisely shown that T-duality relates the AdS$_5$ near-horizon \eqref{D4NS5solutionAdS5} to AdS$_5\times S^5/\mathbb{Z}_k$ vacua in Type IIB, with $k=Q_{\mathrm{NS}5}$ (see \cite{Lozano:2016kum} for more details).
In these vacua, the internal manifold can be locally written as a foliation of three-spheres as it follows
\begin{equation}
\begin{split}\label{orbifoldS5}
ds^2_{S^5/\mathbb{Z}_k}&=d\alpha^2+\sin^2\!\alpha\,d\phi^2+\cos^2\!\alpha\,ds^2_{S^3/\mathbb{Z}_k}\,,\\
d s^2_{S^3/\mathbb{Z}_k}&=\frac14\left[\left(\frac{2d\psi}{k} +\omega \right)^2+ds^2_{S^2}  \right]\,,
\end{split}
\end{equation}
with $d\omega=\text{vol}_{S^2}$. T-duality holds also outside of the near-horizon, meaning that the D4-NS5 system of Table \ref{TypeIIAbranes} can be mapped into a stack of D3 branes with KK monopoles, which are the objects responsible for the orbifolding of the five-sphere.

We can construct the T-dual setup from scratch, directly in Type IIB. To this aim, let us take the system depicted in Table \ref{KKD3branes}.
\begin{table}[http!]
\renewcommand{\arraystretch}{1}
\begin{center}
\scalebox{1}[1]{
\begin{tabular}{c |c cc  c|| c  c  |c c c c}
 branes & $t$ & $x^1$ & $x^2$ & $x^3$ & $y$ & $z$ & $\psi$ & $r$ & $\varphi^1$ & $\varphi^2$ \\
\hline \hline
$\mrm{D}3$ & $\times$ & $\times$ & $\times$ & $\times$ & $-$ & $-$ & $-$ & $-$ & $-$ & $-$ \\
$\mrm{KK}$ & $\times$ & $\times$ & $\times$ & $\times$ & $\times$ & $\times$ & $\mrm{ISO}$ & $-$ & $-$ & $-$ \\
\end{tabular}
}
\caption{The Type IIB brane picture of D4-NS5 intersections of Table \ref{TypeIIAbranes}. T-duality is performed in the $\psi$ direction. The near-horizon limit is described by AdS$_5\times S^5/\mathbb{Z}_k$ geometries.} \label{KKD3branes}
\end{center}
\end{table}
The 10D metric has the following form
\begin{equation}
\label{KKD3branes-solution}
d s_{10}^2 = H_{3}^{-1/2}  ds^2_{\mathbb{R}^{1,3}} + H_{3}^{1/2} \bigl(dy^2+dz^2\bigr)+H_{3}^{1/2}\bigl(H_{\text{KK}}^{-1} (d\psi+2^{-1}Q_{\text{KK}}\omega)^2+H_{\text{KK}}\bigl(dr^2+r^2ds^2_{S^2}  \bigr) \bigr),
\end{equation}
where $Q_{\text{KK}}$ is the KK monopole charge. We may choose the charge distributions as $H_{\text{KK}}=H_{\text{KK}}(r)$ and $H_{3}=H_{3}(y,z,r)$.
As for the D4-NS5 intersection, we then take the semi-localized solution\footnote{We fix the integration constants as in \cite{Lozano:2021fkk}. With this choice of $Q_3$, one precisely obtains $k=Q_5$ via T-duality.} \cite{Youm:1999ti}
\begin{equation}\label{semilocalizedD3KK}
 H_{3}=1+\frac{4\pi Q_{3}Q_{\text{KK}}}{(y^2+z^2+2 Q_{\text{KK}}r)^2}\qquad \text{and}\qquad H_{\text{KK}}=\frac{Q_{\text{KK}}}{2r}\,,
\end{equation}
and cast it into the new parametrization \cite{Oz:1999qd,Cvetic:2000cj}
\begin{equation}\label{AdS5coordIIB}
  z= \mu \sin\alpha\,\sin\phi\,,\qquad\quad y= \mu \sin \alpha\,\cos\phi\,,\qquad\quad r= 4^{-1}\,Q_{\text{KK}}^{-1}\, \mu^2\cos^2\!\alpha\,.
\end{equation}
In these new coordinates the D3-KK background \eqref{semilocalizedD3KK} can be rewritten as \cite{Cvetic:2000cj}
\begin{equation}\label{D3KKmucoord}
 d s_{10}^2 = H_{3}^{-1/2}  ds^2_{\mathbb{R}^{1,3}} + H_{3}^{1/2} \left(d\mu^2+\mu^2 ds^2_{ S^5/\mathbb{Z}_k}  \right) \qquad \text{with}\qquad  H_{3}=1+\frac{4\pi Q_{3}Q_{\text{KK}}}{\mu^4}\,.
\end{equation}
 The $S^5/\mathbb{Z}_k$ metric is precisely the same obtained with T-duality, given in \eqref{orbifoldS5}, with $k=Q_{\text{KK}}$. The near-horizon limit is defined as $\mu\rightarrow 0$ and leads to the geometry \cite{Cvetic:2000cj}
\begin{equation}
\label{KKD3-NHlimit}
\begin{split}
d s_{10}^2 &= L_5^2\,\bigl( ds^2_{\text{AdS}_5}+ds^2_{S^5/\mathbb{Z}_k}\bigr)\,,\\
F_{(5)}&= 4L_5^4 \left(1+\star_{(10)}\right) \vol_{\text{AdS}_5}\,,
\end{split}
\end{equation}
with $ds^2_{\text{AdS}_5}=(4\pi Q_{3}Q_{\text{KK}})^{-1}\,\mu^2ds^2_{\mathbb{R}^{1,3}}+\frac{d\mu^2}{\mu^2}$ and $L_5=(4\pi Q_{3}Q_{\text{KK}})^{1/4}$. For clarity, in the above formulas we also included the five-form flux. The above vacua are BPS$/2$ in Type IIB, namely they preserve 16 real supercharges.
We point out that the case with $k=1$ is special. In this case, supersymmetry is enhanced and we obtain the maximally supersymmetric AdS$_5\times S^5$ vacua dual to $\ma N=4$ SYM theory. We then observe that the presence of KK monopoles affects the vacuum only globally \cite{Cvetic:2000cj}. In other words, to obtain the $\ma N=2$ orbifold vacua \eqref{KKD3-NHlimit} from supergravity one just has to keep the same local geometry of AdS$_5\times S^5$, write the $S^5$ as a foliation as \eqref{orbifoldS5}, and then make the substitution $S^3\rightarrow S^3/\mathbb{Z}_k$.

\section{The supergravity setup}\label{sugrasetup}

Type IIB AdS$_5$ vacua and their fluctuations are very naturally described in five-dimensional supergravity. In this regard, there are many examples in the literature on consistent truncations of Type IIB to five dimensions. In this paper we consider a very simple realization arising from the KK reduction of Type IIB over a $S^5$, namely {\itshape minimal} $\ma N=4$ $\SU2\times \U1$ gauged supergravity. The adjective minimal means that the theory is composed only by the supergravity multiplet. This theory was originally studied by Romans in~\cite{Romans:1985ps} and was obtained as a consistent truncation in~\cite{Lu:1999bw}. Alternatively, this supergravity model can be consistently obtained truncating away all the vector multiplets in the general $\ma N=4$ supergravity \cite{DallAgata:2001wgl}.
In this section, we review the main features of this 5D theory and we discuss the Type IIB reduction leading to it.

\subsection{The 5D Romans supergravity}

The field content of the theory comprises only the supergravity multiplet, whose bosonic part includes \cite{Romans:1985ps}: the 5D gravitational field $g_{\mu\nu}$, three $\SU{2}$ gauge vectors $A_\mu^i$, with $i=1,2,3$, one $\U{1}$ gauge vector $\mathcal{A}_\mu$, two two-forms $B^\alpha_{\mu\nu}$, with $\alpha=1,2$, and a real scalar $X$. Here, $A^i$ transform in the adjoint representation of $\SU{2}$, while $B^\alpha$ transform in the real two-dimensional vector representation of $\SO{2}\cong\U{1}$.

The global symmetry group of the theory is given by $\mathbb{R}\times SO(5)$, where $SO(5)\simeq \text{USp}(4)_R$ is the $R$-symmetry group.
The bosonic part of the action reads~\cite{Lu:1999bw} (see also~\cite{Gauntlett:2007sm})
\begin{equation} \label{action}
  \begin{split}
    \mathcal{S} &= \frac{1}{16\pi G^{(5)}_\mathrm{N}} \int \Bigl[ ( R-\mathcal{V}) \star\!1 - 3X^{-2} \, d X \wedge \star\,d X - \frac12 X^{-2} F^i \wedge \star F^i - \frac12 X^4 \mathcal{F} \wedge \star\mathcal{F} \\
    & - \frac12 X^{-2} B^\alpha \wedge \star B^\alpha + \frac{1}{2g_1} \, \varepsilon_{\alpha\beta} B^\alpha \wedge d B^\beta - \frac12 B^\alpha \wedge B^\alpha \wedge \mathcal{A} - \frac12 F^i \wedge F^i \wedge \mathcal{A} \Bigr] \,,
  \end{split}
\end{equation}
where the scalar can be also parametrized as $X = e^{-\frac{1}{\sqrt6}\phi}$. The fields strengths are given by
\begin{equation}
	F^i = d A^i + \frac12 \, g_2 \varepsilon_{ijk} A^j \wedge A^k  \qquad \text{and}\qquad   \mathcal{F} = d\mathcal{A} \,.
\end{equation}
The real constants $g_1$ and $g_2$ are two gauge parameters associated to the gauging of the subgroup $\SU2\times \U1\subset SO(5)$.
The gauging produces the scalar potential
\begin{equation} \label{scalar-potential}
	\mathcal{V} = -2g_2 \bigl( g_2 X^2 + 2\sqrt2 g_1 X^{-1} \bigr) \,.
\end{equation}
The two-forms can be conveniently packed into one single complex two-form $B \equiv B^1 + i B^2$, with associated field strength
\begin{equation}\label{5Dflux}
	G = d B - i g_1 \mathcal{A} \wedge B \,.
\end{equation}
The complex notation allows us to rewrite the second line of action~\eqref{action} in a more compact form
\begin{equation}
  \begin{split}
    \mathcal{S} &= \frac{1}{16\pi G^{(5)}_\mathrm{N}} \int \Bigl[ ( \ldots ) - \frac12 X^{-2} B \wedge \star \bar{B} + \frac{i}{2g_1} B \wedge \bar{G} - \frac12 F^i \wedge F^i \wedge \mathcal{A} \Bigr] \,.
  \end{split}
\end{equation}
In total, we can identify three inequivalent theories: for $g_2 = 0$ (and $g_1 \neq 0$), $g_1 g_2 > 0$ or $g_1 g_2 < 0$. These are denoted as $\mathcal{N}=4^0$, $\mathcal{N}=4^+$ or $\mathcal{N}=4^-$, respectively~\cite{Romans:1985ps}. We shall focus on the case $\ma N=4^+$. In this case the theory has a maximally supersymmetric AdS vacuum
\begin{equation}\label{AdS5vacua}
 \text{AdS}_5 \qquad \Longrightarrow \qquad X=1\,,
\end{equation}
and for vanishing one- and two-form fields.
Finally, we point out that the above expression is written\footnote{For generic values of $g_1$ and $g_2$, the vacuum is located at $X=\frac{2^{1/6}g_1^{1/3}}{g_2^{1/3}}$.} imposing $g_1=g$ and $g_2 = \sqrt2\,g$ without loss of generality.

We can derive the equations of motion from the action~\eqref{action}. Einstein equations are given by
\begin{equation} \label{einstein-eq}
	\begin{split}
		R_{\mu\nu} &= 3X^{-2} \partial_\mu X \, \partial_\nu X + \frac12 X^{-2} \Bigl( F^i_{\mu\lambda} F_\nu^{i\,\lambda} - \frac16 g_{\mu\nu} F^i_{\rho\sigma} F^{i\,\rho\sigma} \Bigr) + \frac12 X^4 \Bigl( \mathcal{F}_{\mu\lambda} \mathcal{F}_\nu^{\ \lambda} - \frac16 g_{\mu\nu} \mathcal{F}_{\rho\sigma} \mathcal{F}^{\rho\sigma} \Bigr) \\
		& + \frac12 X^{-2} \Bigl( B_{(\mu|\lambda|} \bar{B}_{\nu)}^{\ \lambda} - \frac16 g_{\mu\nu} B_{\rho\sigma} \bar{B}^{\rho\sigma} \Bigr) + \frac13 g_{\mu\nu} \mathcal{V} \,.
	\end{split}
\end{equation}
The field equations of vectors and two-form fields are\footnote{We point out that the first equation has a different sign with respect to~\cite{Lu:1999bw}. This is consistent with our conventions on the action and the Hodge dual.}
\begin{equation} \label{maxwell-eq}
	\begin{split}
		d\bigl( X^{-2} \star\!F^i \bigr) &= -g_2 X^{-2} \varepsilon_{ijk} A^j \wedge \star F^k - F^i \wedge \mathcal{F} \,, \\
		d\bigl( X^4 \star\!\mathcal{F} \bigr) &= -\frac12 F^i \wedge F^i - \frac12 B \wedge \bar{B} \,, \\
		X^2 \star\!G &= -i g_1 B \,.
	\end{split}
\end{equation}
Finally, the equation of motion of the scalar field has the following form
\begin{equation} \label{scalar-eq}
	\begin{split}
		d\bigl(X^{-1} \star d X) &= -\frac16 X^{-2} F^i \wedge \star F^i + \frac13 X^4 \mathcal{F} \wedge \star\mathcal{F} - \frac16 X^{-2} B \wedge \star \bar{B} \\
		& - \frac23 g_2 \bigl( g_2 X^2 - \sqrt2 g_1 X^{-1} \bigr) \star\!1 \,.
	\end{split}
\end{equation}

\subsubsection{The SUSY variations}

Let's now study the SUSY variations of fermions. Our theory is five-dimensional and with $\ma N=4$ supersymmetry, which implies that we have four gravitini $\psi_{\mu x}$ and four gaugini $\chi_x$, with $x=1,\ldots,4$. These spinors transform in the fundamental representation of the $R$-symmetry group $\USp{4}_R$.

The components of the SUSY parameters $\epsilon_x$ are organized in symplectic-Majorana spinors. We refer to Appendix \ref{app:gamma-matr} for information relating to the spinorial structures involved in our work. The action~\eqref{action} is invariant under the following supersymmetry transformations~\cite{Romans:1985ps}%
\footnote{In order to obtain these expressions we need to divide by 2 all the fields in~\cite{Romans:1985ps}, multiply by 2 the gauge couplings and change the signature of the metric, including the redefinition $\gamma^a \mapsto i\gamma^a$.}
\begin{equation} \label{susy-var}
	\begin{split}
		\delta\psi_{\mu x} &= \mathcal{D}_\mu \epsilon_x - i \gamma_\mu \ma{T}_x^{\ y} \epsilon_y - \frac{i}{12\sqrt2} \bigl( \gamma_\mu^{\ \nu\rho} - 4\delta^\nu_\mu \gamma^\rho \bigr) \biggl( \ma{S}_{\nu\rho x}^{\quad\,y} + \frac{1}{\sqrt2} \, \ma{S}_{\nu\rho x}'^{\quad\,y} \biggr) \epsilon_y \,, \\
		\delta\chi_x &= \frac{i}{2\sqrt2} (\partial_\mu \phi) \gamma^\mu \epsilon_x + \ma{R}_x^{\ y} \epsilon_y + \frac{1}{4\sqrt6} \gamma^{\mu\nu} \bigl( \ma{S}_{\mu\nu x}^{\quad\,y} - \sqrt2 \, \ma{S}_{\mu\nu x}'^{\quad\,y} \bigr) \epsilon_y \,,
	\end{split}
\end{equation}
where the covariant derivative reads
\begin{equation}
	\mathcal{D}_\mu \epsilon_x = \partial_\mu \epsilon_x + \frac14 \, \omega_\mu^{\ ab} \gamma_{ab} \epsilon_x + \frac{g_1}{2} \mathcal{A}_\mu (\Gamma_{45})_x^{\ y} \epsilon_y + \frac{g_2}{2} A^i_\mu (\Gamma_{i45})_x^{\ y} \epsilon_y \,,
\end{equation}
and we defined the following quantities
\begin{align} \label{susy-TRS}
	& \ma{T}_x^{\ y} = \biggl( \frac{1}{3\sqrt2} g_2 X + \frac16 g_1 X^{-2} \biggr) (\Gamma_{45})_x^{\ y} \,,  \qquad
	&& \ma{R}_x^{\ y} = \biggl( \frac{1}{\sqrt6} g_2 X - \frac{1}{\sqrt3} g_1 X^{-2} \biggr) (\Gamma_{45})_x^{\ y} \,, \nonumber \\
	& \ma{S}_{\mu\nu x}^{\quad\,y} = X^{-1} \bigl[ F^i_{\mu\nu} (\Gamma_i)_x^{\ y} + B^\alpha_{\mu\nu} (\Gamma_\alpha)_x^{\ y} \bigr] \,,  \qquad
	&& \ma{S}_{\mu\nu x}'^{\quad\,y} = X^2 \mathcal{F}_{\mu\nu} \delta^y_x \,.
\end{align}
We note that the above relations explicitly depend on two types of gamma matrices: $\{\Gamma_{I}\}$ with $I=(i,\alpha)$ associated with the $R$-symmetry group $\USp{4}_R\simeq\Spin5_R$, and the 5D spacetime matrices $\{\gamma_{\mu}\}$. Again, we refer to Appendix \ref{app:gamma-matr} for more details.

Importantly, from the above variations one can extract a superpotential $\ma W$ of the following form
\begin{equation}\label{superpotential}
	\mathcal{W}(X) = \frac{1}{3\sqrt2} g_2 X + \frac16 g_1 X^{-2}\,.
\end{equation}
Specifically, the scalar potential~\eqref{scalar-potential} and the coefficients~\eqref{susy-TRS} can be written as
\begin{equation}
	\mathcal{V} = 12 \bigl[ X^2 (\partial_X \mathcal{W})^2 - 4\mathcal{W}^2 \bigr] \,,  \qquad
	\ma{T}_x^{\ y} = \mathcal{W} \, (\Gamma_{45})_x^{\ y} \,,  \qquad
	\ma{R}_x^{\ y} = \sqrt3 X \, \partial_X \mathcal{W} \, (\Gamma_{45})_x^{\ y} \,,
\end{equation}
using the explicit expression of the superpotential \eqref{superpotential}.

\subsection{The dimensional reduction from Type IIB}\label{IIBreduction}

We now discuss how 5D Romans supergravity can be obtained from consistent truncation of Type IIB over a $S^5$. This reduction was originally obtained in~\cite{Lu:1999bw} through a prescription on the internal geometry preserving $\SU2\times \U1$ symmetry. The general idea is to write the $S^5$ as a foliation of a $S^3$ and a circle, precisely as we wrote in \eqref{orbifoldS5}. Away from the vacuum, the 5D fields describe the deformations of metric and fluxes. In what follows, we will present the KK reduction restricted to vanishing one-form fields (the gauge fields $A^i$ and $\ma A$ defined in previous section) and refer to~\cite{Lu:1999bw} for the general truncation.
The reduction Ansatz of the 10D metric is given by~\cite{Lu:1999bw}
\begin{equation}
 \begin{split}\label{reductionMetric}
  &ds^2_{10}=\Delta^{1/2}ds^2_5+g^{-2}\Delta^{-1/2}ds_{M_5}\,,\\
  &ds^2_{M_5}=X\Delta\, d\alpha^2+X^2\sin^2\!\alpha\, d\psi^2+X^{-1}\cos^2\!\alpha\, ds^2_{S^3}\,,
 \end{split}
\end{equation}
where $\Delta=X^{-2}\sin^2\!\alpha+X\cos^2\!\alpha$. The coordinate $\psi$ parameterizes a circle as in \eqref{orbifoldS5}. A peculiar feature of this truncation is that the 10D scalar degrees of freedom are constant, namely $\Phi=0$ and $C_0=0$.
Introducing the notation $s=\sin\alpha$, $c=\cos\alpha$, we can write the Ansatz for gauge fluxes as it follows~\cite{Lu:1999bw}
\begin{equation}
\begin{split}\label{reductionFluxes}
 &G_{(3)}=-2^{-1/2}g^{-1}e^{-i\psi}\,\left(c\,d\alpha-is \,d\psi\right)\wedge B-2^{-1/2}g^{-1}se^{-i\psi}dB\,,\\
& F_{(5)}=(1+\star_{(10)})\left[2g\, (X^2c^2+X^{-1}s^2+X^{-1})\,\vol_{5} -3g^{-1}sc X^{-1}\star_{(5)}dX\wedge d\alpha \right]\,,\\
 \end{split}
\end{equation}
where $G_{(3)}=F_{(3)}+iH_{(3)}$ is the 10D complex three-flux in Type IIB. The five-flux explicitly wraps on $\vol_{5}$, which is the volume form of the 5D spacetime.

In the above relations, all the dynamical quantities appearing on the right-hand side of the equalities are defined over the 5D spacetime. On the left-side are the Type IIB metric and fluxes involved in the truncation.
If one substitutes the above prescription in the equations of motion of Type IIB supergravity, one obtains the field equations of 5D Romans' supergravity \eqref{einstein-eq}--\eqref{scalar-eq}, with $g_2=\sqrt 2 g_1$ and $g_1=g$.

As a comment that will be useful later, we note that in this dimensional reduction the vacuum fluctuations from the two-form $B$ are responsible for a non-trivial profile of the three-fluxes in 10D, while the scalar $X$ controls the deformations of the internal space and of the five-form flux.
From the above formulas we observe that for $X=1$ and $B=0$ we reproduce the supersymmetric AdS$_5\times S^5$ vacua \eqref{KKD3-NHlimit} with radius $L_5=g^{-1}$.

Finally, we point out that thanks to the formulas above, one can also use Romans supergravity to describe fluctuations within AdS$_5\times S^5/\mathbb{Z}_k$ vacua. In fact, as we discussed in section \ref{ref:D3KKsystem}, the inclusion of KK monopoles in the D3 branes stack does not modify the local structure of the vacuum. Instead, they affect the 10D spacetimes only globally, through the orbifolding $S^3\rightarrow S^3/\mathbb{Z}_k$ within the $S^5$ geometry.

\section{New AdS$_2$ solutions with two-form potentials}\label{solutions}

In this section we study AdS$_2\times S^2\times I$ bulk geometries in 5D Romans supergravity. We explicitly solve the BPS equations in two situations: when the AdS$_2$ and $S^2$ warp factors scale in the same way, namely they have equal radii, and when they scale differently. In both systems we are able to find solutions preserving half supersymmetries (BPS$/2$ in 5D), featured by two-form gauge potentials and the scalar field with a non-trivial radial profile. In the case where AdS$_2$ and $S^2$ have the same radius we obtain a domain wall-like solution interpolating between an asymptotically locally AdS$_5$ geometry and a spacetime singularity. In the more complicated situation where AdS$_2$ and $S^2$ scale differently, we find a regular solution interpolating between locally AdS$_5$ and an AdS$_2\times \mathbb{R}^3$ geometry.

\subsection{The Ansatz for bosonic fields}

In this paper we consider half-BPS five-dimensional backgrounds described by the following metrics
\begin{equation}\label{generalMetricAnsatz}
d s_5^2 = e^{2U(\rho)} \, d s_{\AdS_2}^2 + d\rho^2 + e^{2W(\rho)} \, d s_{S^2}^2\,,
\end{equation}
where $d s_{\AdS_2}^2$ and $ds_{S^2}^2$ denote the unit radius metrics on $\AdS_2$ and $S^2$, respectively. Moreover, we suppose that the scalar field has the following non-trivial profile
\begin{equation}
 X=X(\rho)\,.
\end{equation}
As far as the gauge sector is concerned, we take two two-forms of the type
\begin{equation}
 B^\alpha = b_\alpha(\rho) \, \vol_{\AdS_2} + c_\alpha(\rho) \, \vol_{S^2}\,,
\end{equation}
and vanishing vectors $A^i=\ma A=0$.

We point out that the above Ansatz for the two-forms is over-determined. In fact, from the equations of motion~\eqref{maxwell-eq} we obtain the two relations
\begin{equation} \label{b2-c1}
	b_2 = k \, b_1 \,,  \qquad\quad  c_1 = -k \, c_2 \,,
\end{equation}
where $k$ is a real constant. The remaining functions $b_1$ and $c_2$ must satisfy
\begin{equation} \label{Max}
	b_1' = -g_1 e^{2U-2W}X^{-2} c_2 \,,  \qquad\quad     c_2' = -g_1 e^{2W-2U} X^{-2}b_1 \,.
\end{equation}
Given the conditions~\eqref{b2-c1}, we can now show that it is always possible to rotate the fields $B^\alpha$ to set $b_2$ and $c_1$ to zero.
As we mentioned in previous section, the action~\eqref{action} and the equations of motion \eqref{einstein-eq}--\eqref{scalar-eq} are invariant under an $\SO{2}$ rotation of the doublet $B^\alpha$, namely
\begin{equation}
	\begin{pmatrix}
		B^1 \\
		B^2
	\end{pmatrix} \mapsto
	\begin{pmatrix}
		\cos\xi & -\sin\xi \\
		\sin\xi & \cos\xi
	\end{pmatrix}
	\begin{pmatrix}
		B^1 \\
		B^2
	\end{pmatrix}
	\qquad  \iff  \qquad  B \mapsto e^{i\xi} B \,,
\end{equation}
where $\xi$ is the parameter of the rotation.
Also the supersymmetry variations \eqref{susy-var} are invariant if we, additionally, rotate the doublet of gamma matrices $(\Gamma_4,\Gamma_5)$. A fundamental fact is that this rotation does not affect the combination $\Gamma_{45}$, which remains unaltered.
If we now take $\xi = -\arctan(k)$ we obtain
\begin{equation}
	\begin{pmatrix}
		\cos\xi & -\sin\xi \\
		\sin\xi & \cos\xi
	\end{pmatrix}
	\begin{pmatrix}
		b_1 \, \vol_{\AdS_2} - k\,c_2 \, \vol_{S^2} \\
		k\,b_1 \, \vol_{\AdS_2} + c_2 \, \vol_{S^2}
	\end{pmatrix}
	= \sqrt{1+k^2}
	\begin{pmatrix}
		b_1 \, \vol_{\AdS_2} \\
		c_2 \, \vol_{S^2}
	\end{pmatrix} \,.
\end{equation}
In what follows, we shall employ this symmetry to set $k=0$. 
 With this choice the two-forms are given by
\begin{equation}
 B^1= b(\rho) \, \vol_{\AdS_2} \qquad\quad \text{and}\qquad\quad B^2= c(\rho) \, \vol_{S^2} \,,
\end{equation}
where we introduced the notation $b_1=b$ and $c_2=c$.

\subsection{The Ansatz for the Killing spinor}

Let's now consider the prescription for the Killing spinor. As we mentioned previously, the SUSY parameters $\epsilon_x$ are 5D symplectic-Majorana spinors, which we will take of the following general form\footnote{We refer to Appendix \ref{app:gamma-matr} for our conventions on gamma matrices and index notations.}
\begin{equation}\label{generalKillingSpinor}
	\epsilon_x = Y(\rho) \bigl[ \cos\bigl(\theta(\rho)\bigr) \, \mathbb{I}_4 \delta_x^y + \sin\bigl(\theta(\rho)\bigr) \, \gamma_{01}  (\Gamma_5)_x^{\ y} \bigr] \epsilon^0_y \,,
\end{equation}
where $Y(\rho)$ and $\theta(\rho)$ are two real functions. We recall that the matrices $(\Gamma_I)_x^{\,\,y}$, with $I=1,\dots,5$, are the gamma matrices of the $R$-symmetry group $\USp4_R\simeq \Spin5_R$. The spinor $\epsilon^0_x$ can be decomposed as follows
\begin{equation}\label{generalKillingSpinor1}
  \epsilon^0_x = e^{i \varphi\gamma_2} \,\vartheta_x \otimes \zeta_x \,,
\end{equation}
where $\varphi$ is a phase fixed by the Killing spinor equations.
Here, $\vartheta_x=\vartheta_x(x^{\hat{\mu}})$ are four Majorana Killing spinors on $\AdS_2$, while $\zeta_x=\zeta_x(x^{\tilde{\mu}})$ is a set four symplectic-Majorana Killing spinors on $S^2$. Explicitly, these two-dimensional spinors satisfy the following relations
\begin{equation}\label{2DkillingEq}
	\vartheta_x^* = \mathcal{B}_2^\beta \vartheta_x \,,  \quad  \hat{\nabla}_{\hat{\mu}} \vartheta_x = \frac{\kappa_x}{2} \beta_{\hat{\mu}} \vartheta_x \qquad  \text{and}  \qquad
	\zeta_x^* = \Omega^{xy} \mathcal{B}_2^\rho \zeta_y \,,  \quad  \tilde{\nabla}_{\tilde{\mu}} \zeta_x = \frac{\kappa'_x}{2} \rho_* \rho_{\tilde{\mu}} \zeta_x \,,
\end{equation}
where the 2D gamma matrices $\beta_{\hat \mu}$ and $\rho_{\tilde \mu}$ are defined as in \eqref{gamma-deco}.
The first and the third relations are respectively the Majorana and symplectic-Majorana conditions for 2D Lorentzian and Euclidean spinors. The matrices $\ma B_2^\beta$ and $\ma B_2^\rho$ are introduced in \eqref{Bdecomposition}. The second and four equations are the Killing spinor equations for spinors on AdS$_2$ and $S^2$ in $D=2$. These equations are defined up to the signs $\kappa_x,\kappa'_x=\pm1$.

In this work, we study solutions breaking half of the supersymmetries of the AdS$_5$ vacuum. To this aim, we will impose the projection condition
\begin{equation} \label{projection}
	\gamma_2 (\Gamma_{45})_x^{\ y} \epsilon^0_y = i\epsilon^0_x \,.
\end{equation}
As we are going to see, this condition, together with the above prescriptions for bosonic fields and the Killing spinor, reproduces a consistent set of BPS equations.

\subsection{The singular $\mathrm{AdS}_2\times S^2$ domain wall}\label{subsec:chargedDW}

We start our analysis of BPS solutions by imposing $U=W$ in the 5D metric \eqref{generalMetricAnsatz}. This is equivalent to consider the following Ansatz on bosonic fields
\begin{equation}
\begin{split}
 ds_5^2 &= e^{2U(\rho)}\,\left(ds^2_{\text{AdS}_2}+ds^2_{S^2}  \right)+d\rho^2\,,\\
 B &= b(\rho)\,\text{vol}_{\text{AdS}_2}+ i\,c(\rho)\,\text{vol}_{S^2}\,,\\
 X &= X(\rho)\,.
 \end{split}
\end{equation}
The quantities $b$ and $c$ in the complex two-form are related each other. This can be seen by plugging the above prescription into Maxwell equations~\eqref{Max}, leading to $c^2 = b^2 + \text{constant}$. In what follows, we will then focus on backgrounds such that
\begin{equation}
 b(\rho)=c(\rho)\,.
\end{equation}
We can now evaluate the SUSY variations \eqref{susy-var} over this field configuration. First, one can see that the Killing spinor \eqref{generalKillingSpinor} gets strongly simplified with the condition $U=W$, namely
\begin{equation}
 \theta(\rho)=0\,,
\end{equation}
together with the condition $\kappa'_x=\kappa_x$ on the signs of the 2D Killing equations \eqref{2DkillingEq}. The Killing spinor boils down to the form
\begin{equation}
  \epsilon_x =Y\, e^{i \varphi\gamma_2} \,\vartheta_x \otimes \zeta_x \,.
\end{equation}

Imposing the projection \eqref{projection}, from the condition of vanishing SUSY variations we extract the following BPS equations
\begin{equation} \label{f=g_first-order}
	U' = -2\mathcal{W} \,,  \qquad    X' = 2X^2 \, \partial_X \mathcal{W}\,, \qquad
	b' = -\frac{g \,b}{X^2} \,,  \qquad  Y' = -Y \,\mathcal{W} \,,
\end{equation}
where the superpotential $\ma W$ was given in \eqref{superpotential}.
The phase $\varphi$ is fixed by the following condition
\begin{equation} \label{phase-constr}
	e^{-2i\varphi \gamma_2} \epsilon_x + \frac{\kappa_x b\,e^{-U}}{\sqrt2 X}  (\Gamma_5)_x^{\ y} \epsilon_y = 0 \,.
\end{equation}
In the explicit representation~\eqref{Gamma-rep} the matrix $\Gamma_5$ is diagonal. In this basis the above equations decouple and can be solved independently, leading to the following conditions
\begin{equation} \label{phase+b1}
	\varphi = 0 \,,  \qquad\quad  b  = \kappa \, \sqrt2 X e^U\,,  \qquad\quad  \kappa_1 = \kappa_4 = -\kappa_2 = -\kappa_3 \equiv \kappa \,.
\end{equation}
It can be shown that the expression of $b$ is consistent with the first-order equation for $b'$ in~\eqref{f=g_first-order}.
We can now integrate equations~\eqref{f=g_first-order} introducing the following reparametrization
\begin{equation}
	d\rho = e^{V(\mu)} \, d \mu  \qquad\quad  \text{with}  \qquad\quad  e^{-V} = 2X^2 \, \partial_X \ma{W} \,.
\end{equation}
Expressed in this coordinate, the BPS equation for the scalar takes the form $X'=1$. We can choose the particular solution $X=\mu$.
The remaining BPS equations become easily solvable, leading to the solution
\begin{equation}\label{chargedDW}
	e^{2U} = \frac{\mu}{g^2(1-\mu^3)} \,,  \qquad e^{2V}=\frac{9\mu^2}{4g^2(1-\mu^3)^2}\,,\qquad
	b= \frac{\sqrt2 \, \mu^{3/2}}{g(1-\mu^3)^{1/2}} \,,\qquad   Y = e^{U/2}\,,
\end{equation}
where we set $\kappa=1$, $g_2 = \sqrt2 g$ and $g_1=g$, and fixed the integration constant in $e^U$ to a convenient value.
The solution is a BPS$/2$ domain wall defined along the $\mu$ coordinate. As it is manifest from the non-trivial profile for $b$, this domain wall is charged under the two-forms $B^1$ and $B^2$, respectively wrapping the AdS$_2$ and $S^2$ directions.

One can explicitly verify that a curvature singularity lies at $\mu=0$, while in $\mu=1$ the metric presents a conformal boundary, therefore we take the coordinate $\mu$ to range in the interval $\mu\in[0,1]$. At the conformal boundary, the solution is asymptotically locally AdS$_5$. This can be observed by expanding the Ricci scalar in a neighborhood of $\mu=1$, obtaining
\begin{equation}\label{AdS5Ricci}
 R=-20g^2+\ma O(\mu-1)^2\,,\qquad \qquad X=1+\ma O(\mu-1)\,.
\end{equation}
These are precisely the Ricci scalar and the VEV of the field $X$ at the AdS$_5$ vacuum. We stress that the vacuum geometry is reproduced only locally. This can be explicitly observed noticing that the two-form $B$ is non-zero in the asymptotics, thus breaking the isometries of the AdS$_5$ vacuum.

\subsection{A new Janus in AdS$_5$}
\label{subsec:janus-sol}

We now focus on five-dimensional geometries in which we allow the AdS$_2$ and $S^2$ factors to scale independently. Specifically, we consider the following prescription for the fields,
\begin{equation}
	\begin{split}	\label{JanusAnsatz}
		ds_5^2 &= e^{2U(\rho)}\,ds^2_{\text{AdS}_2}+e^{2W(\rho)}\,ds^2_{S^2}+d\rho^2\,,\\
		B &= b(\rho)\,\text{vol}_{\text{AdS}_2}+ i\,c(\rho)\,\text{vol}_{S^2}\,,\\
		X &= X(\rho)\,.
	\end{split}
\end{equation}
In order to obtain a consistent set of BPS equation, we need to employ the general Ansatz for the Killing spinor \eqref{generalKillingSpinor}, that we recall for the reader's convenience:
\begin{equation}\label{generalKillingSpinorJanus}
  \epsilon_x = Y(\rho) \bigl[ \cos\bigl(\theta(\rho)\bigr) \, \mathbb{I}_4 \delta_x^y + \sin\bigl(\theta(\rho)\bigr) \, \gamma_{01}  (\Gamma_5)_x^{\ y} \bigr] \epsilon^0_y \, ,
\end{equation}
where $\epsilon^0_x$ is given in \eqref{generalKillingSpinor1}.
We stress that in this more general situation the Killing spinor depends on two dynamical variables, $Y(\rho)$ and $\theta(\rho)$. This is coherent with the fact that we are considering two variables $U$ and $W$ for variations of AdS$_2$ and $S^2$ factors. This present situation is, of course, more involved than the case $U=W$ studied in previous section.

Given the above prescriptions, we are able to find a consistent set of BPS equations imposing one projection condition, which is again the condition given in \eqref{projection}. It follows that our solutions will be half-BPS in 5D, precisely as the previous case with $U=W$. Evaluating the SUSY variations over the Ansatz \eqref{JanusAnsatz} and specifying a Killing spinor of the form \eqref{generalKillingSpinorJanus}, we obtain the following set of BPS equations
\begin{equation}
	\begin{split} \label{BPS}
	  U' &= \frac{1}{3\cos(2\theta)} \left(  -2(2+\cos(4\theta)) \mathcal{W} - 4 \sin^2(2\theta) X \, \partial_X \mathcal{W} + \kappa \, e^{-U} \sin(2\theta) \right) \,, \\
	  W' &= \frac{1}{3\cos(2\theta)} \left( (-7+\cos(4\theta)) \mathcal{W} + 2\sin^2(2\theta) X \, \partial_X \mathcal{W} + \kappa \, e^{-U} \sin(2\theta) \right) \,, \\
	  X' &= \frac{X}{3 \cos(2\theta)} \left( (5+\cos(4\theta)) X \, \partial_X \mathcal{W} + 2\sin^2(2\theta) \mathcal{W} - \kappa \, e^{-U} \sin(2\theta) \right) \,, \\
	  Y' &= \frac{\Xi}{6\cos(2\theta)} \left(  -2(2+\cos(4\theta)) \mathcal{W} -4 \sin^2(2\theta) X \, \partial_X \mathcal{W}  + \kappa \, e^{-U} \sin(2\theta) \right) \,,\\
	  \theta' &= \sin(2\theta) \bigl( \mathcal{W} - X \, \partial_X \mathcal{W} \bigr)\,,
  \end{split}
\end{equation}
where $\ma W$ is the superpotential \eqref{superpotential}. In these equations we have the freedom to choose two signs, namely $\kappa=\pm1$ and $\kappa'=\pm1$. These are defined as in~\eqref{phase+b1} in terms of the $\kappa_x$ and $\kappa_x'$ appearing in the 2D Killing spinor equations \eqref{2DkillingEq}.

In addition to the first-order equations, we find three algebraic constraints between the dynamical variables. Two of them provide the two-form variables $b$ and $c$ in terms of the other fields,
\begin{equation}
\begin{split}\label{b1c2}
  &b = 2\sqrt2 \kappa X e^U - \sqrt2 X e^{2U} \left( \kappa' \, e^{-W} \cos(2\theta) + 6\mathcal{W} \sin(2\theta) \right) \,, \\
  &c =  2\sqrt2 \kappa' X e^W - \frac{\sqrt2 X e^{2W}}{\cos(2\theta)} \left( \kappa \, e^{-U} - 6\mathcal{W} \sin(2\theta) \right) \,.
  \end{split}
\end{equation}
The third constraint is a consistency condition relating the dynamical flows of the AdS$_2$ and $S^2$ metric factors to the other quantities, namely the scalar $X$ and the spinor variable $\theta$. The condition has the following form
\begin{equation}\label{algConst}
  -\frac{\kappa}{2} + \frac{\kappa'\,e^{U-W}}{2} \cos(2\theta) + e^U \bigl( 2\mathcal{W} + X \, \partial_X \mathcal{W} \bigr) \sin(2\theta) = 0 \,.
\end{equation}

In order to integrate the BPS equations, we will follow the same strategy employed in \cite{Conti:2024rwd}.
First, from \eqref{BPS} we immediately observe that the standard relation $Y=e^{U/2}$ for the overall spinor variable $Y$ is still valid for this family of backgrounds. Then, as a crucial observation, we point out that from the constraint \eqref{algConst} we can extract two integrals of motion
\begin{equation} \label{C1C2}
  C_1=e^{U-W} \cos(2\theta)\,,  \qquad   \qquad  C_2=e^U \sin(2\theta) \bigl( 2\mathcal{W} + X \, \partial_X \mathcal{W} \bigr) \,.
\end{equation}
One can explicitly verify that $C_1$ and $C_2$ are constant along the flow by taking their $\rho$-derivatives and imposing the BPS equations \eqref{BPS}. The algebraic constraint \eqref{algConst} can be then written as
\begin{equation} \label{alg:constrC}
	-\frac{\kappa}{2} + \frac{\kappa'\,C_1}{2} + C_2 = 0 \,.
\end{equation}
At this point the strategy is to relate all the quantities to $X$ and $\theta$ using \eqref{C1C2}. Directly from \eqref{C1C2} we immediately obtain the expressions for $U$ and $W$,
\begin{equation} \label{fhC1C2}
	e^U = \frac{C_2}{g X} \sin^{-1}(2\theta)\,,  \qquad\quad  e^W = \frac{C_2}{g\,C_1 X}\,\cot(2\theta) \,.
\end{equation}
We can now use the equation for $\theta$ to trade the $\rho$ coordinate for $\theta$ itself as
\begin{equation}
	d\rho = \frac{2X^2}{g} \sin^{-1}(2\theta)\,d\theta \,.
\end{equation}
We are ready to solve the equation for $X$ in \eqref{BPS}. Using the explicit form of the superpotential \eqref{superpotential} and the expression for $U$ in \eqref{fhC1C2}, the equation for $X$ can be exactly integrated, giving the following solution
\begin{equation}\label{XSOL}
  X = \frac{C_2^{1/3} \cos^{1/3}(2\theta) \sin^{-2/3}(2\theta)}{\left( \sqrt2 C_2 C_3 + (C_2-\kappa) \log(\cot\theta) + C_2 \sin^{-2}(2\theta)\cos(2\theta) \right)^{1/3}} \, ,
\end{equation}
where $C_3$ is an integration constant and $\theta \in(0, \frac{\pi}{4})$.
The expressions for $U$, $W$, $b$ and $c$ can be obtained plugging \eqref{XSOL} in \eqref{fhC1C2} and \eqref{b1c2}. Even if this solution seems to reproduce a very intricate spacetime dependence, in what follows we will show that the 5D geometry can be written in a very simple form.

\subsubsection{The solution and its asymptotics}

Our solution \eqref{XSOL} is defined over the interval $\theta \in (0, \frac{\pi}{4})$. Let's consider the behavior near $\theta=\frac{\pi}{4}$. In this limit the scalar field behaves as
\begin{equation}
  X^{-1} = \frac{C_3^{1/3}}{2^{1/6}\left(\frac{\pi}{4}-\theta\right)^{1/3}} + \frac{2^{1/3} (2 C_2 - \kappa)}{3 C_2 C_3^{2/3}} \left( \theta -\frac{\pi}{4} \right)^{2/3} + \mathcal{O}\left(\theta-\frac{\pi}{4} \right)^{4/3}.
\end{equation}
Importantly, we observe that for $C_3 = 0$ the scalar field becomes regular at this point,
\begin{equation}
  X \sim \frac{C_2^{1/3}}{( 2 C_2 - \kappa )^{1/3}} \equiv X_0 \, .
\end{equation}
We can now show that in $\theta =\frac{\pi}{4}$ the metric describes a regular $\AdS_2 \times \mathbb{R}^3$ geometry. To this aim, we point out that the constraint \eqref{alg:constrC} can be rewritten as
\begin{equation}
	 2C_2 -\kappa = -\kappa'\,C_1  \qquad  \implies  \qquad  C_2^2 X_0^{-6} = C_1^2 \,,
\end{equation}
where, we recall, $\kappa$ and $\kappa'$ can be $\pm 1$.
If we expand the metric around $\theta = \frac{\pi}{4}$ and use this condition we obtain
\begin{equation}
  d s_5^2 \sim  \frac{C_2^2}{g^2 X_0^2} d s^2_{\AdS_2} + \frac{4X_0^4}{g^2} \left( d\theta^2 + \left( \theta - \frac{\pi}{4} \right)^2  d s^2_{S^2} \right) ,
\end{equation}
where the second term is precisely the three-dimensional Euclidean space $\RR^3$ in spherical coordinates.

Given this regular behavior in the $\theta \rightarrow \frac{\pi}{4}$ limit, let us now consider the full bulk solution. Introducing the auxiliary parameter
\begin{equation}
  \lambda = 1 - \frac{\kappa}{C_2}
\end{equation}
and using the constraint \eqref{alg:constrC}, we can write the two constants $C_1$ and $C_2$ as
\begin{equation}
  C_1 = - \frac{\kappa(1+\lambda)}{\kappa'(1-\lambda)} \, , \qquad C_2 = \frac{\kappa}{1-\lambda} \, .
\end{equation}
The full solution can then be written as a family of AdS$_2\times S^2\times I$ geometries defined by $\lambda \in (-1,1)$. Their general form is given by
\begin{equation}\label{5DJanus}
  \begin{split}
    d s^2_5 & = \frac{1}{g^2 X^2 \sin^2(2\theta)} \left( \frac{1}{(1-\lambda)^2} d s^2_{\AdS_2} + 4X^6 d\theta^2 + \frac{\cos^2(2\theta)}{(1+\lambda)^2} d s^2_{S^2} \right) \, ,\\
    X & =\left(1+ \lambda \, \frac{\sin^2(2\theta)}{\cos(2\theta)} \log(\cot\theta) \right)^{-1/3}\, ,\\
    B & = - \frac{\sqrt2 \lambda}{g(1-\lambda)^2} \sin^{-1}(2\theta) \left[ 1+  \frac{\sin^2(2\theta)}{\cos(2\theta)} \log(\cot\theta)  \right] \vol_{\AdS_2} \\
	&  - i\,\frac{\sqrt2 \lambda}{g(1+\lambda)^2} \cot(2\theta) \left[ 1 - \frac{\sin^2(2\theta)}{\cos(2\theta)} \log(\cot\theta)  \right] \vol_{S^2} \, ,
  \end{split}
\end{equation}
where we wrote the two-form fields in the complex notation.

We immediately observe that for $\lambda=0$ we obtain the global $\AdS_5$ vacuum with radius $L_{5}^2 = g^{-2}$, parameterized as
\begin{equation}
  d s^2_5 = \frac{1}{g^2 \sin^2(2\theta)} \left( d s^2_{\AdS_2}  + 4 d\theta^2 +  \cos^2(2\theta) d s^2_{S^2} \right) \, .
\end{equation}
For $\lambda \neq 0$ the solution is locally AdS$_5$ in the $\theta \rightarrow 0$ limit. In fact at the leading order in $\theta \rightarrow 0$, the solution \eqref{5DJanus} can be written as
\begin{equation}
\begin{split}
  d s^2_5 &\sim \frac{1}{4g^2 \theta^2} \left( \frac{1}{(1-\lambda)^2}\,d s^2_{\AdS_2}  + 4 d\theta^2 +  \frac{1}{(1+\lambda)^2}\, d s^2_{S^2} \right) \, ,\\
  X&\sim 1\,,\\
  B &\sim-\frac{\lambda}{\sqrt2 g\,(1-\lambda)^2\,\theta}\,\text{vol}_{\text{AdS}_2}-i\frac{\lambda}{\sqrt2g\,(1+\lambda)^2\theta}\,\text{vol}_{S^2}\,.
  \end{split}
\end{equation}
In this limit the Ricci scalar reproduces the AdS$_5$ behavior \eqref{AdS5Ricci}, namely $R\sim -20 g^2$. The contribution of the two-form to the equations of motion is then subleading with respect to the metric, but still non-zero. For this reason the AdS$_5$ isometries are irremediably broken and the geometry is only locally AdS$_5$ at $\theta=0$.

\section{The Type IIB origin}\label{IIBorigin}

In this section we discuss the interpretation of our 5D solutions in Type IIB. To this aim, we provide the uplift to ten dimensions of both types of AdS$_2\times S^2\times I$ solutions constructed in the previous section, by making use of formulas of section \ref{IIBreduction}.

As far as the singular solution \eqref{chargedDW} is regarded, we provide the exact brane interpretation as the near-horizon of D1-F1-D5-NS5 defect branes ending on the D3-KK system of Table \ref{KKD3branes}.
 This brane setup was introduced in \cite{Lozano:2021fkk}, where it was explicitly shown that the near-horizon geometry is a class of AdS$_2\times S^2\times S^1\times S^3/\mathbb{Z}_k$ solutions fibered over two intervals. Here we provide the change of coordinates in 10D connecting our singular domain wall to that near-horizon geometry.

 A crucial property of the brane solution presented in \cite{Lozano:2021fkk} is the {\itshape smearing} of D1-F1-D5-NS5 branes along the transverse directions of D3 branes. This property allows to decouple the field equations of the two groups of branes. This implies that the entire backreaction of D1-F1-D5-NS5 branes is breaking up the worldvolume isometries of D3 branes to the curved geometry
	\begin{equation}
	 \mathbb{R}^{1,3}\longrightarrow \text{AdS}_2\times S^2\,,
	\end{equation}
where, importantly, AdS$_2$ and $S^2$ have the same radius. As it is argued for similar supergravity setups \cite{Conti:2024rwd}, this property seems to be intimately related to the singular behavior of the domain wall \eqref{chargedDW}.

As we showed in section \ref{ref:D3KKsystem}, D3 branes with KK monopoles are associated to AdS$_5\times S^5/\mathbb{Z}_k$ vacua, defining the ambient four-dimensional SCFT. The intersection D1-F1-D5-NS5 ends on the worldvolume of the D3s, breaking its conformal isometries through a spacetime dependent deformation. The result is a superconformal quantum mechanics describing a line defect within the 4D SCFT \cite{Lozano:2021fkk}.

Unfortunately, we are not able to provide a precise brane interpretation for our regular Janus solutions \eqref{5DJanus}. Nevertheless we point out that the Type IIB uplift is still described by solutions of the type AdS$_2\times S^2\times S^1\times S^3/\mathbb{Z}_k$. The main difference now is that the singularity is smoothed out by allowing  AdS$_2$ and $S^2$ to have different radii. This seems to suggest that the brane setup is still the same described above, but this time with D1-F1-D5-NS5 branes considered {\itshape fully-localized} in their transverse space (see \cite{Bachas:2013vza,Dibitetto:2020bsh,Conti:2024rwd} for a similar discussion for AdS$_3\times S^3$ backgrounds in M-theory).

\subsection{The D1-F1-D5-NS5-KK-D3 system}

Let's start with the Type IIB origin of the singular domain wall \eqref{chargedDW}. To this aim, we first compute the uplift using  formulas of section~\ref{IIBreduction}. From equations \eqref{reductionMetric} and \eqref{reductionFluxes}, we obtain the following 10D geometry
\begin{equation}\label{chargedDWUplift}
 \begin{split}
  ds^2_{10}&=\frac{\mu\Delta^{1/2}}{g^2 (1-\mu^3)} \bigl( ds^2_{\text{AdS}_2} + ds^2_{S^2} \bigr) + \frac{9\mu^2\Delta^{1/2}}{4g^2 (1-\mu^3)^2}\,d\mu^2 \\
  &+g^{-2}\Delta^{-1/2}\left[\mu\Delta\, d\alpha^2+\mu^2\sin^2\!\alpha\, d\psi^2+\mu^{-1}\cos^2\!\alpha\, ds^2_{S^3}\right]\,,
 \end{split}
\end{equation}
with $\Delta=\mu^{-2}\sin^2\!\alpha+\mu\cos^2\!\alpha$. This metric describes an AdS$_2\times S^2\times S^1\times S^3$ spacetime fibered over two intervals parameterized by the coordinates $(\mu,\alpha)$. The corresponding fluxes are given by
\begin{equation}
 \begin{split}
	G_{(3)}&=-\frac{\mu^{3/2}\,e^{-i\psi}}{g^2 (1-\mu^3)^{1/2}}\,\left[c\, d\alpha-i s\, d\psi+\frac{3s}{2\mu(1-\mu^3)}\,d\mu \right]\wedge \left(\text{vol}_{\AdS_2}+i\,\text{vol}_{S^2} \right),\\
	F_{(5)}&=(1+\star_{(10)})\left[\frac{3\mu^2(\mu^3c^2+s^2+1)}{g^4 (1-\mu^3)^3}\, d\mu -\frac{2sc}{g^4 (1-\mu^3)}\, d\alpha\right]\wedge\vol_{\AdS_2}\wedge \vol_{S^2}\,,
 \end{split}
\end{equation}
where we used the notation $s=\sin\alpha$ and $c=\cos\alpha$. At the extrema of $\mu \in [0, 1]$ the solution reproduces the AdS$_5\times S^5$ vacuum and a curvature singularity.
Following what observed in section \ref{IIBreduction}, by topological substitution $S^3\rightarrow S^3/\mathbb{Z}_k$, we can generate AdS$_2\times S^2\times S^1\times S^3/\mathbb{Z}_k$ solutions with AdS$_5\times S^5/\mathbb{Z}_k$ asymptotics. We stress that there is no supersymmetry enhancement in the asymptotic limit because the AdS$_5$ vacua geometry are realized only locally, and not globally.

Let's now discuss the D-brane interpretation of this Type IIB solution. In what follows, we show that it is related by a coordinates change to the near-horizon geometry of a precise brane system. The idea is to consider a stack of D3s with KK monopoles (see section \ref{ref:D3KKsystem}) and to intersect it with a suitable bound state of D1-F1-D5-NS5 branes. This intersection was introduced in \cite{Lozano:2021fkk} and is depicted in Table \ref{DefectBraneSystem}. Let's summarize the main features of this brane setup.
\begin{table}[http!]
\renewcommand{\arraystretch}{1}
\begin{center}
\scalebox{1}[1]{
\begin{tabular}{c  |c cc  c|| c  c  c c c c}
 branes & $t$ & $u$ & $\theta^1$ & $\theta^2$ & $y$ & $z$ & $\psi$ & $r$ & $\varphi^1$ & $\varphi^2$ \\
\hline \hline
$\mrm{D}3$ & $\times$ & $\times$ & $\times$ & $\times$ & $-$ & $-$ & $-$ & $-$ & $-$ & $-$ \\
$\mrm{KK}$ & $\times$ & $\times$ & $\times$ & $\times$ & $\times$ & $\times$ & $\mrm{ISO}$ & $-$ & $-$ & $-$ \\
\hline
$\mrm{D}1$ & $\times$ & $-$ & $-$ & $-$ & $\times$ & $-$ & $-$ & $-$ & $-$ & $-$ \\
$\mrm{F}1$ & $\times$ & $-$ & $-$ & $-$ & $-$ & $\times$ & $-$ & $-$ & $-$ & $-$ \\
$\mrm{D}5$ & $\times$ & $-$ & $-$ & $-$ & $\times$ & $-$ & $\times$ & $\times$ & $\times$ & $\times$ \\
$\mrm{NS}5$ & $\times$ & $-$ & $-$ & $-$ & $-$ & $\times$ & $\times$ & $\times$ & $\times$ & $\times$ \\
\end{tabular}
}
\caption{Brane system from \cite{Lozano:2021fkk} describing D1-F1-D5-NS5 branes ending on D3-KK branes. Our 5D domain wall \eqref{chargedDW} describes the near-horizon limit of this system when D1-F1-D5-NS5 charges are {\itshape smeared} along the internal directions of the vacuum.} \label{DefectBraneSystem}
\end{center}
\end{table}
We may start writing the general metric of the intersection
\begin{equation}
\label{General10DmetricDefect}
\begin{split}
d s_{10}^2 &= H_{3}^{-1/2}  \left[- H_{1}^{-1/2} H_{\mathrm{F}1}^{-1} H_{5}^{-1/2}dt^2 +H_{1}^{1/2}  H_{5}^{1/2}H_{\mathrm{NS}5}\left(du^2+u^2ds^2_{S^2} \right)  \right] \\
&+ H_{3}^{1/2} \left[H_{1}^{-1/2}  H_{5}^{-1/2}H_{\mathrm{NS}5}dy^2+H_{1}^{1/2} H_{\mathrm{F}1}^{-1} H_{5}^{1/2}dz^2\right]\\
&+H_{3}^{1/2}H_{1}^{1/2}  H_{5}^{-1/2}\,\left[H_{\text{KK}}^{-1}\left(d\psi+2^{-1}Q_{\text{KK}}\omega \right)^2+H_{\text{KK}}\left(dr^2+r^2ds^2_{\tilde S^2}  \right) \right]\,,
\end{split}
\end{equation}
where for the D3-KK system we use the same notations\footnote{Here we denote with $\tilde S^2$, parameterized by $(\varphi^1, \varphi^2)$, the two-sphere transverse to D3 branes, in order to distinguish it from the $S^2$ parameterized by $(\theta^1, \theta^2$), that turns out to belong to the 5D domain wall.} of section \ref{ref:D3KKsystem}. Specifically, we consider $H_3=H_3(y,z,r)$ and $H_{\text{KK}}=H_{\text{KK}}(r)$. We remind again that we can turn off the KK monopoles reproducing supersymmetry enhancement and a round $S^3$.

Let's now consider the charge distributions of D1-F1-D5-NS5 branes. We point out that their presence breaks the isometries on the D3 worldvolume. To keep this into account, we introduced the coordinates $(u, \theta^1, \theta^2)$ on the space $\mathbb{R}_u^3$ where D1-F1-D5-NS5 branes are localized. We then suppose that they are smeared along the directions $(y, z, \psi, \mathbb{R}^3_r)$. In other words, we require that D1-F1-D5-NS5 branes are entirely localized within the worldvolume of the D3s. Concretely, we assume that $H_{1}(u)$, $H_{\mathrm{F}1}(u)$, $H_{\mathrm{NS}5}(u)$, $H_{5}(u)$. From this assumption it follows that the equations of motion for the D3-KK system and those for D1-F1-D5-NS5 branes decouple \cite{Lozano:2021fkk}. Specifically, one gets the harmonic functions
\begin{equation}\label{solutionD1-F1-D5-NS5}
  H_{1}=H_{\text{NS}5}=1+\frac{Q_{1}}{u} \qquad \text{and} \qquad H_{\mathrm{F}1}=H_5=1+\frac{Q_{\mathrm{F}1}}{u}\,,
\end{equation}
where $Q_{1}$ and $Q_{\mathrm{F}1}$ are parameters in relation to the quantized charges of the corresponding branes. We observe that we can introduce a near-horizon limit $u\rightarrow 0$, corresponding to zooming in on D1-F1-D5-NS5 branes. The above geometry takes the following form  \cite{Lozano:2021fkk}
\begin{equation}
\begin{split}\label{nearHorizonTypeIIB}
 ds_{10}^2 &= Q_{1}^{3/2}Q_{\mathrm{F}1}^{1/2} H_{3}^{-1/2}\bigl(ds^2_{\text{AdS}_2}+ds^2_{S^2}\bigr) +Q_{1}^{1/2}Q_{\mathrm{F}1}^{-1/2} H_{3}^{1/2} \bigl( dy^2+dz^2 \bigr)\\
&+Q_{1}^{1/2}Q_{\mathrm{F}1}^{-1/2}H_{3}^{1/2}\bigl(H_{\text{KK}}^{-1}(d\psi+2^{-1}Q_{\text{KK}}\omega)^2+H_{\text{KK}}\bigl(dr^2+r^2ds^2_{\tilde S^2}  \bigr) \bigr)\,.
\end{split}
\end{equation}
We thus observe that, in the $u\rightarrow 0$ limit, the unique effect of these branes is to deform the D3 worldvolume to AdS$_2\times S^2$. This is particularly evident by comparing this solution with the geometry \eqref{KKD3branes-solution} of the pure D3-KK system. We point out that this fact is due to our assumption on the smearing of D1-F1-D5-NS5, that also implies that the transverse space of D3s remains the same after intersecting D1-F1-D5-NS5 branes.

We can now specify this backgrounds to the semi-localized solution \eqref{semilocalizedD3KK} for D3-KK branes and perform the change of coordinate \eqref{AdS5coordIIB}. The solution \eqref{nearHorizonTypeIIB} becomes \cite{Lozano:2021fkk}
\begin{equation}
\begin{split}
ds_{10}^2 &= Q_{1}^{3/2}Q_{\mathrm{F}1}^{1/2}    H_{3}^{-1/2}\left(ds^2_{\text{AdS}_2}+ds^2_{S^2}\right) +Q_{1}^{1/2}Q_{\mathrm{F}1}^{-1/2} H_{3}^{1/2} \left(d\mu^2+\mu^2 ds^2_{S^5/\mathbb{Z}_k}  \right)\,,\\
 H_{3}&=1+\frac{4\pi Q_{\mathrm{KK}} Q_{3}}{\mu^4}\,,
\end{split}
\end{equation}
where the $S^5/\mathbb{Z}_k$ metric has the form \eqref{orbifoldS5}. Following the same logic of section \ref{ref:D3KKsystem}, we can finally take the $\mu \rightarrow 0$ limit obtaining a locally AdS$_5\times S^5/\mathbb{Z}_k$ geometry. This analysis is performed explicitly in \cite{Lozano:2021fkk}, including the derivation of fluxes.

In what follows we show that our 5D singular domain wall can be precisely mapped in \eqref{nearHorizonTypeIIB}. If we go back to the Type IIB uplift \eqref{chargedDWUplift} and we compare it with the near-horizon geometry \eqref{nearHorizonTypeIIB}, we obtain the following relations\footnote{
In order to perform the mapping precisely, we need to rescale all the fields in~\cite{Lozano:2021fkk} in order to set $\Phi=0$, as required in~\cite{Lu:1999bw}. To do this, we exploit the following symmetry of Type~IIB supergravity
\begin{equation} \label{scaling-symm}
	d\hat{s}_\mathrm{s.f.}^2 = \lambda^2 d s_\mathrm{s.f.}^2 \,,  \qquad  e^{\hat{\Phi}} = \lambda^2 e^\Phi \,,  \qquad
	\hat{B}_{(2)} = \lambda^2 B_{(2)} \,,  \qquad	 \hat{F}_{(n)} = \lambda^{n-3} F_{(n)} \,,
\end{equation}
with $n=1,3,5$ and $\lambda$ a strictly positive constant.}
\begin{equation}
	r = \frac{e^{2U} c^2 X^{-1}}{4Q_\mathrm{F1}^2 Q_\mathrm{KK} g^2} \,, \qquad  \qquad
	z + i y = \frac{i \, e^{i\psi} s \, b}{\sqrt2\, g\, Q_\mathrm{F1}} \,,
\end{equation}
where on the right-hand side we have the 5D quantities $e^{2U}$, $X$ and $b$ defined by our solution \eqref{chargedDW}. In order to realize the mapping one also needs to impose the identification
\begin{equation}\label{H3change}
	H_3 = \frac{Q_1 Q_\mathrm{F1}^3}{e^{4U} \Delta} \,.
\end{equation}
	Notice that the above expression for $H_3$ is equivalent to the semi-localized solution for D3-KK system\footnote{The equation of motion for $H_3$, written in (3.11) of \cite{Lozano:2021fkk} is satisfied by \eqref{H3change} once we impose the explicit 5D solution.} \eqref{semilocalizedD3KK} once we impose the explicit 5D solution \eqref{chargedDW}.

	We then showed that the 5D singular domain wall \eqref{chargedDW} describes the physics of D1-F1-D5-NS5 branes intersecting the D3-KK bound state, studied in \cite{Lozano:2021fkk}.

\subsection{The 10D Janus}

Let's compute the uplift to Type IIB of the regular Janus solution \eqref{5DJanus}. Using formulas \eqref{reductionMetric} we obtain the following geometry
\begin{equation}\label{JanusUplift}
 \begin{split}
  ds^2_{10}&=\frac{\Delta^{1/2}}{g^2X^2\sin^{2}(2\theta)}\,\left[(1-\lambda)^{-2}ds^2_{\text{AdS}_2}+4X^6d\theta^2+(1+\lambda)^{-2}\cos^2(2\theta)ds^2_{S^2} \right] \\
  &+g^{-2}\Delta^{-1/2}\left[X\Delta\, d\alpha^2+X^2\sin^2\!\alpha\, d\psi^2+X^{-1}\cos^2\!\alpha\, ds^2_{S^3}\right]\,,\\
   X_{\,\,} &=\left(1+ \lambda \, \frac{\sin^2(2\theta)}{\cos(2\theta)} \log(\cot\theta) \right)^{-1/3} \,,
 \end{split}
\end{equation}
with $\Delta=X^{-2}\sin^2\!\alpha+X\cos^2\!\alpha$. This metric describes a one-parameter class of AdS$_2\times S^2\times S^1\times S^3$ backgrounds fibered over two intervals, parameterized by the coordinates $(\theta, \alpha)$. As in the previous case, the round $S^3$ can be substituted by the orbifold $S^3/\mathbb{Z}_k$. The 10D fluxes have the general form of the uplift \eqref{reductionFluxes},
\begin{equation}
	\begin{split}
	G_{(3)}&=-2^{-1/2}g^{-1}e^{-i\psi}\,\left(c\,d\alpha-is \,d\psi\right)\wedge B-2^{-1/2}g^{-1}se^{-i\psi}dB\,,\\
	\tilde F_{(5)}&=\frac{\cot^{2}(2\theta)}{2g^4(1-\lambda^2)^2X^7}\left[\frac{8\ma F_5 X^5}{\sin^3(2\theta)}\,d\theta-\frac{3csX'}{\sin(2\theta)}d\alpha  \right]\wedge \text{vol}_{\text{AdS}_2}\wedge\text{vol}_{S^2}\,,
	\end{split}
\end{equation}
where we employed the notations $s=\sin\alpha$, $c=\cos\alpha$, $\ma F_5=(X^2c^2+X^{-1}s^2+X^{-1})$ and $F_{(5)}=(1+\star_{(10)})\tilde F_{(5)}$. The complex three-flux $G_{(3)}$ explicitly depends on the 5D two-form \eqref{5DJanus} that we may recall here,
\begin{equation}
\begin{split}
B &= - \frac{\sqrt2 \lambda}{g(1-\lambda)^2} \sin^{-1}(2\theta) \left[ 1+  \frac{\sin^2(2\theta)}{\cos(2\theta)} \log(\cot\theta)  \right] \vol_{\AdS_2} \\
	& - i\,\frac{\sqrt2 \lambda}{g(1+\lambda)^2} \cot(2\theta) \left[ 1 - \frac{\sin^2(2\theta)}{\cos(2\theta)} \log(\cot\theta)  \right] \vol_{S^2} \, .
\end{split}
\end{equation}
First, we point out that for $\lambda=0$ we recover the global AdS$_5\times S^5/\mathbb{Z}_k$ vacua, while for $\lambda \neq 0$ the uplifted solution describes a 10D Janus solution interpolating between $\theta=0$ and $\theta=\frac{\pi}{4}$. Applying the results obtained in five dimensions, we recover locally AdS$_5\times S^5/\mathbb{Z}_k$ vacua for $\theta \rightarrow 0$ and a regular background AdS$_2\times \mathbb{R}^3\times S^1\times S^3/\mathbb{Z}_k$ fibered over the interval in the limit $\theta\rightarrow\frac{\pi}{4}$.

Unfortunately, we are not able to find a brane solution whose near-horizon limit is equivalent to \eqref{JanusUplift}. Nevertheless, we observe that the amount of supersymmetry preserved and the topology, AdS$_2\times S^2\times S^1\times S^3/\mathbb{Z}_k$, are the same as the singular solution \eqref{chargedDWUplift}. This leads us to suppose that our two 5D solutions may have a similar brane origin in Type IIB. The crucial difference with respect to the singular case is that the AdS$_2$ and $S^2$ factors scale differently, allowing the $S^2$ to shrink smoothly in $\theta=\frac{\pi}{4}$. These facts suggest the possibility that the brane setup underlying our Janus solution is still represented by the D1-F1-D5-NS5-KK-D3 intersection depicted in Table \ref{DefectBraneSystem} and by a general 10D metric of the type \eqref{General10DmetricDefect}. In order to obtain a near-horizon geometry with a different scaling between AdS$_2$ and $S^2$ variations, one should modify the prescription over the charge distribution of defect branes D1-F1-D5-NS5. For instance, with the more general prescription
\begin{equation}\label{fullylocalized}
	 H_1=H_1(u, z, r)\,,\qquad H_{\text{F}1}=H_{\text{F}1}(u, y, r)\,,\qquad H_{5}=H_{5}(u, z)\,,\qquad H_{\text{NS}5}=H_{\text{NS}5}(u, y)\,,
\end{equation}
defect brane charges are fully-localized, namely they are localized within the space $\mathbb{R}^3_u$ and along the transverse directions of D3 branes $(y, z, \mathbb{R}^3_r)$.
Given this more general assumption, the technical challenge is providing a particular solution for \eqref{fullylocalized} that admit AdS$_2$ and AdS$_5$ limits in a suitable combination of coordinates $(u, y, z, r)$. In this case in fact, the field equations for defect and mother branes do not separate anymore (see the discussion after \eqref{General10DmetricDefect}), implying that we cannot take independently the two near-horizon limits.

\section{Holographic calculations}
\label{sec:holo-ren}

In this section we perform the holographic computation of the on-shell action of the Janus solution constructed in section~\ref{subsec:janus-sol} and we study the one-point functions of the holographic field theory.
The methods used in this section follow the prescriptions of the \emph{holographic renormalization}~\cite{Henningson:1998gx,deHaro:2000vlm}. Before starting with our analysis, let's remind the logic underlying this holographic procedure.

Our Janus solution is characterized by a locally $\AdS_5$ asymptotic geometry, therefore we claim that a four-dimensional SCFT is defined at the boundary.
As usual, the presence of non-vanishing fields in the bulk reflects in deformations in the boundary CFT, obtained by adding relevant operators and, possibly, turning on vacuum expectation values (VEVs) for these. The AdS/CFT dictionary \cite{Maldacena:1997re, Witten:1998qj} states a one-to-one correspondence between operators on the boundary and fields in the bulk, identifying the sources of the former with the value on the boundary of the latter.
Additionally, the string theory partition function, as a functional of the boundary values~$\phi_0$, equals the CFT generating function of correlators. In the saddle-point approximation, this relation becomes $W_\mathrm{CFT}[\phi_0] = -\mathcal{S}_\mathrm{on\mhyphen shell}[\phi_0]$, where $W_\mathrm{CFT}[\phi_0]$ is the generator of connected correlation functions and $\mathcal{S}_\mathrm{on\mhyphen shell}[\phi_0]$ is the supergravity on-shell action. It follows that the on-shell action can be used as a functional generator for the correlators of the CFT.

In quantum field theory correlation functions suffer from UV divergences, which are mirrored by the IR (\ie\ near-boundary) divergences appearing on the gravity side. This divergent behavior is caused by two different contributions: first, the bulk action diverges due to the infinite volume of the spacetime $M$ and, second, the Gibbons--Hawking--York term is ill-defined because the induced metric on $\partial M$ diverges.
In order to circumvent this issue and to obtain a finite on-shell action one needs to systematically remove all the divergences. Such a result can be achieved applying the holographic renormalization procedure~\cite{Henningson:1998gx,deHaro:2000vlm}.

This approach starts with a regularization prescription realized imposing a cutoff in the spacetime, integrating the bulk action up to the cutoff, and evaluating the corresponding boundary terms. This computation, which is performed in the near-boundary region, relies on the series expansion in the cutoff parameter of the metric and the additional fields. The leading contributions play the role of sources in the dual field theory.
The regularized action contains a finite set of terms that diverge when the cutoff is removed. Taking care of these infinities is the purpose of the second part of the procedure, the renormalization. As it happens in quantum field theory, we can cancel the divergences order-by-order introducing appropriate local covariant counterterms, which, in our case, can be entirely expressed in terms of the induced boundary metric and fields, evaluated at the cutoff~\cite{deHaro:2000vlm}.
Once the counterterms are added to the regularized action,%
\footnote{One can always decide to include additional finite counterterms, which translates in choosing a different renormalization scheme.}
the cutoff can be removed and the resulting renormalized action turns out to be finite.

By construction, the renormalized on-shell action is invariant under bulk diffeomorphisms, except for the ones that generate Weyl transformations of the boundary metric.%
\footnote{The CFT on the boundary is endowed with a conformal class of metrics $[g_0]$, but the renormalization procedure dictates to select a representative $g_0$, thus breaking explicitly the conformal invariance.}
Conformal invariance of the field theory defined on the boundary can possibly be broken, thus leading to the emergence of \emph{conformal anomalies}. These anomalies are closely related to logarithmically divergent terms in the action and, for this reason, appear only when the boundary is even-dimensional~\cite{Henningson:1998gx,deHaro:2000vlm}.

Following the logic outlined above, we devote the first part of this section to the analysis of the near-boundary behavior of our Janus solution \eqref{5DJanus}. In the second part, we explicitly compute the renormalized on-shell action. Finally, in the last part we make some considerations regarding the conformal weight of the CFT dual operators and calculate their one-point function.

\subsection{The boundary expansion}

As we mentioned in the opening of this section, the first ingredients required for our computation are the expansions of the fields in a region close to the boundary. Since the Janus solution \eqref{5DJanus} is asymptotically locally $\AdS_5$, we can employ the Fefferman--Graham (FG) expansion to study the 5D metric near the boundary. In general, the FG expansion near the boundary has the form
\begin{equation} \label{FG-deco}
	ds_5^2 = \frac{1}{z^2} \bigl( dz^2 + g_{ij}(z,x) \, dx^i dx^j \bigr) \,,
\end{equation}
where the boundary is located at $z=0$ and the metric tensor $g_{ij}$ admits the following series expansion in powers of~$z$
\begin{equation} \label{g-deco}
	g(z,x) = g_0(x) + z^2 g_2(x) + z^4 \big[ g_4(x) + \log z \, \tilde{g}_4(x) + \log^2\!z \, \hat{g}_4(x) \bigr] + \mathcal{O}(z^5) \,.
\end{equation}
The terms involving odd power of $z$ are expected to vanish up to order $z^4$, while the logarithmic contributions at $z^4$ are required because the boundary is even-dimensional~\cite{Henningson:1998gx} (see also~\cite{Skenderis:2002wp}).
We expect for the scalar field $X$ and the two-forms $B^\alpha$ the following asymptotic expansions
\begin{equation} \label{XB-deco}
	\begin{split}
		X(z,x) &= 1 + z^2 \bigl[ X_2(x) + \log z \, \tilde{X}_2(x) \bigr] + \mathcal{O}(z^3) \,, \\
		B^\alpha(z,x) &= z^{-1} B^\alpha_{-1}(x) + z \bigl[ B^\alpha_1(x) + \log z \, \tilde{B}^\alpha_1(x) \bigr] + \mathcal{O}(z^2) \,,
	\end{split}
\end{equation}
where $B^\alpha_{-1}$, $B^\alpha_1$ and $\tilde{B}^\alpha_1$ are two-forms living on the boundary.
Although these expansions are not the most general ones (see~\cite{Alday:2014bta} for an example in $D=6$), they still capture the asymptotic behavior of our Janus solution, as will be clear shortly.

We now focus on our explicit solution~\eqref{5DJanus} and put it into the FG form. This computation is crucial to extract from the expansion the various holographic quantities of our solution, such as the anomalies and the one-point functions.
The Fefferman--Graham expansion can be obtained expressing the coordinate $\theta$ in~\eqref{5DJanus} in terms of the FG coordinate $z$. The comparison of the two metrics leads to\footnote{From now on, we will set to one the radius of the $\AdS_5$ vacuum, hence $g=1$.}
\begin{equation}
  \frac{dz}{z} = \frac{2X^2 d\theta}{\sin(2\theta)} \, .
\end{equation}
Since the above equation cannot be solved analytically, we will search for a solution in a neighborhood of the boundary. To this end, we expand the right-hand side around $\theta=0$, integrate term by term and expand again the exponential of the result. In this way we obtain the perturbative expansion
\begin{equation}
  \begin{split}
    z(\theta) & = \theta + \frac{1}{3} \, \theta^3 \bigl( 1 - 2\lambda + 4\lambda \log\theta \bigr) + \frac{1}{90} \, \theta^5 \bigl(12 -20 \lambda + 45 \lambda^2 \\
    & + 60\lambda (2 - 3\lambda)  \log\theta + 280 \lambda ^2 \log^2\!\theta \bigr) + \mathcal{O}(\theta^6) \, ,
  \end{split}
\end{equation}
that can be inverted as
\begin{equation}
  \begin{split}
    \theta(z) &= z - \frac13 \, z^3 \bigl( 1 - 2\lambda + 4\lambda \log z \bigr) + \frac{1}{90} \, z^5 \bigl( 18 - 60\lambda - 5\lambda^2\\
    & + 20\lambda (6-7\lambda) \log z + 200\lambda^2 \log^2\!z \bigr) + \mathcal{O}(z^6) \, .
  \end{split}
\end{equation}
Plugging this expression in the various functions describing our solution~\eqref{5DJanus} we obtain
\begin{align} \label{FG-solution}
  e^{2U} & = \frac{1}{4(1-\lambda)^2 z^2} + \frac{3-2\lambda}{6(1-\lambda)^2} + \frac{[9-3\lambda^2 - 12\lambda(4-\lambda) \log z - 8\lambda^2 \log^2\!z] z^2}{36(1-\lambda)^2} + \mathcal{O}(z^3) \, , \nonumber \\
  e^{2W} & = \frac{1}{4(1+\lambda)^2z^2} - \frac{3+2\lambda}{6(1+\lambda)^2} + \frac{[9-3\lambda^2 + 12\lambda(4+\lambda) \log z - 8\lambda^2 \log^2\!z] z^2}{36(1+\lambda)^2} + \mathcal{O}(z^3) \, , \nonumber \\
  b & = -\frac{\lambda}{\sqrt2 (1-\lambda)^2 z} - \frac{[3-2\lambda - 4(3-\lambda) \log z] \lambda z}{3\sqrt2 (1-\lambda)^2} + \mathcal{O}(z^2) \, ,\\
  c & = -\frac{\lambda}{\sqrt2(1+\lambda)^2 z} + \frac{[3+2\lambda - 4(3+\lambda) \log z] \lambda z}{3\sqrt2 (1+\lambda)^2} + \mathcal{O}(z^2) \, , \nonumber \\
  X & = 1 + \frac{4\lambda}{3} z^2 \log z + \frac{8\lambda^2}{9} z^4 + \mathcal{O}(z^5) \, .\nonumber
\end{align}

Before addressing the renormalization of the on-shell action, we compute the conformal dimensions of the operators corresponding to the scalar $X$ and to the tensor field $B$. These can be obtained from the linearized equations of motion near the AdS$_5$ boundary. By inserting 
$X - 1 \sim z^{\Delta_X}$ into \eqref{scalar-eq} we derive
\begin{equation}
  \Delta_X (\Delta_X-4) = -4 \, ,
\end{equation}
that implies that the dual operator has conformal dimension $\Delta_X = 2$. Thus we expect that the VEV of the operator dual to the scalar field is proportional to the coefficient of the $z^2$ term in the FG expansion of $X$, which agrees with the explicit computation below, resulting in~\eqref{vev_X}.
Analogously, if we plug $B_{ij} \sim  z^{\Delta_B-2}$ in \eqref{os-max} we obtain
\begin{equation}
  (\Delta_B-2)^2 = 1
\end{equation}
and so $\Delta_B = 3$ for the operator dual to $B$. Therefore, the VEV of the operators dual to the two-forms will be related to the linear term in the FG expansion of $B^\alpha$. From equations~\eqref{FG-solution} we see that, for our solution, the operator dual to the scalar has vanishing VEV, whereas are non-zero the ones corresponding to $B$.

\subsection{The on-shell action}

We may start rewriting the gravity action employing the equations of motion. Using the third Maxwell equation and substituting the expression for the Ricci scalar $R$ coming from the Einstein equations, the on-shell bulk action reads
\begin{equation}
	\mathcal{S}_\mathrm{bulk} = \frac{1}{16\pi G^{(5)}_\mathrm{N}} \int_M d^5x \, \sqrt{-G} \, \biggl( \frac23 \, \mathcal{V} + \frac{1}{12} X^{-2} B^\alpha_{\mu\nu} B^{\alpha\mu\nu} \biggr) \,,
\end{equation}
where now we called $G$ the five-dimensional metric to avoid confusion with the near-boundary metric \eqref{g-deco}. In order to reproduce a well-posed variational problem, we need to add the Gibbons--Hawking--York (GHY) term
\begin{equation} \label{GHYterm}
	\mathcal{S}_\mathrm{GHY} = \frac{1}{8\pi G^{(5)}_\mathrm{N}} \int_{\partial M} d^4x \, \sqrt{-h} \, h^{ij} K_{ij} \,.
\end{equation}
Here, $h_{ij}$ is the induced boundary metric and $K_{ij}$ its extrinsic curvature tensor.

Since on-shell action is divergent, we need to add counterterms on the boundary to subtract the divergent terms and renormalize it. The detailed construction is shown in Appendix \ref{app:counterterms}, here we report the total divergence
\begin{equation}
  \label{S:div}
  \begin{split}
    \ma S_{\mathrm{div}} & = \frac{1}{16\pi G^{(5)}_\mathrm{N}} \int d^4 x \, \sqrt{-g_0} \, \biggl\{ 6 \epsilon^{-4} + \log\epsilon \biggl[ \frac{1}{12} R[g_0]^2 - \frac14 R_{ij}[g_0]R^{ij}[g_0]\\
    & - \frac18 \tr\bigl( g_0^{-1} \mathrm{Ric}[g_0] g_0^{-1} B^\alpha_{-1} g_0^{-1} B^\alpha_{-1} \bigr) + 3 \tilde X^2_2 \biggr] \biggr\}
  \end{split}
\end{equation}
and the expression for the total counterterm
\begin{equation}
  \begin{split}\label{S-count}
    \ma S_{\mathrm{c.t.}} & = \frac{1}{16\pi G^{(5)}_\mathrm{N}}  \int_{\partial M} d^4x \, \sqrt{-h} \, \biggl\{ -6 - \frac12 R[h] -6(1-X)^2 - \log\epsilon \biggl[ \frac{1}{12} R[h]^2 \\
    & - \frac14 R_{ij}[h] R^{ij}[h] + \frac{1}{16} \tr\bigl[(h^{-1} B^\alpha)^2 (h^{-1} B^\beta)^2\bigr] \biggr] - \frac{3}{\log\epsilon}(1-X)^2 \biggr\} \, .
  \end{split}
\end{equation}
As we mentioned in the introduction, the logarithmic divergences in the above expression are related to the conformal anomalies of the theory. Specifically, the gravitational anomaly reproduces its standard expression (see \cite{Henningson:1998gx}), while the remaining terms are due to matter.
The scalar anomaly agrees with the result of \cite{Bianchi:2001kw}, while the one coming from the $B$ form, to the best of our knowledge, has never been constructed before. To validate its expression one should compare it with the VEV of the trace of the stress--energy tensor, which captures the whole set of anomalies of the theory.

We can now define the regularized on-shell action by adding the counterterms to the on-shell bulk action integrated for $\theta \geq \epsilon$ and to the boundary term ($\mathcal{S}_{\mathrm{GHY}} + \mathcal{S}_{\mathrm{c.t.}}$) evaluated at $\theta=\epsilon$. It can be shown that the contribution to the bulk action coming from $\theta\rightarrow \frac{\pi}{4}$ is zero, while in the regularized action all the divergences cancel out. We can therefore take the $\epsilon \rightarrow 0$ limit and obtain the renormalized on-shell action
\begin{equation}
  \label{Sren:tot}
  \ma S_{\mathrm{ren}} = - \frac{1}{16\pi G^{(5)}_\mathrm{N}} \cdot \frac{1+4\lambda^2}{4(1-\lambda^2)^2} \mr{Vol}(\AdS_2) \mr{Vol}(S^2) \, .
\end{equation}
When $\lambda$ is set to zero we recover the the on-shell action for the global $\AdS_5$. Subtracting this value from \eqref{Sren:tot} gives the on-shell action for the defect solution \eqref{5DJanus}
\begin{equation}
  \ma S_{\mathrm{ren}}(\mathrm{defect}) = - \frac{1}{16\pi G^{(5)}_\mathrm{N}} \cdot \frac{\lambda^2(6-\lambda^2)}{4(1-\lambda^2)^2} \mr{Vol}(\AdS_2) \mr{Vol}(S^2) \, .
\end{equation}

As a last comment, we stress that all the computations were performed in the Lorentzian signature. The Euclidean on-shell action, which is the one entering in the gravitational path integral, is related by $\mathcal{S}_\mathrm{E} \leftrightarrow -\mathcal{S}_\mathrm{ren}$, where the volume of AdS must be replaced by the volume of its Euclidean counterpart.

\subsection{One-point functions and VEVs}

We now move to the discussion of the correlation functions in the holographic field theory.
We remind that the boundary fields associated to the leading terms in the FG expansion (as $g_0$, $\tilde{X}_2$ and $B_{-1}^\alpha$ in \eqref{FG-deco}, \eqref{XB-deco}) correspond to the sources for the dual operators.
Once we have identified the sources, we can compute the associated one-point functions, that are defined in a standard way~\cite{deHaro:2000vlm, Bianchi:2001kw}. Starting with the scalar $X$ we have
\begin{equation}
	\langle\mathcal{O}_X\rangle = \frac{1}{\sqrt{-g_0}} \frac{\delta \mathcal{S}_\mathrm{ren}}{\delta \tilde{X}_2} =
	\lim_{\epsilon\to0} \biggl( \frac{\log\epsilon}{\epsilon^2} \frac{1}{\sqrt{-h}} \frac{\delta \mathcal{S}_\mathrm{sub}}{\delta X} \biggr) \,,
\end{equation}
where $\mathcal{S}_\mathrm{sub} = \mathcal{S}_\mathrm{reg} + \mathcal{S}_\mathrm{c.t.}$ and $\mathcal{S}_\mathrm{reg}$ is the sum of the bulk action and GHY term computed with the cut-off $\epsilon$.
Explicitly, $\delta \mathcal{S}_\mathrm{GHY} = 0$, while the bulk action gives
\begin{equation}
  \delta\mathcal{S}_{\mathrm{bulk}} = \frac{1}{16\pi G^{(5)}_\mathrm{N}} \int_M d^5x \, \Bigl[ \sqrt{-G} \, \mathcal{E}(X) \, \delta X - \partial_\mu \bigl( \sqrt{-G} \, 6X^{-2} \, \partial^\mu X \, \delta X \bigr) \Bigr] \,,
\end{equation}
where $\mathcal{E}(X)$ is a term proportional to the EOMs for the scalar $X$ and, therefore, vanishes when computed on-shell.
We shall now apply Gauss's theorem
\begin{equation}
	\int_M d^5x \, \sqrt{-G} \, \nabla_\mu V^\mu = \oint_{\partial M} d^4x \, \sqrt{-h} \, n_\mu V^\mu \,,
\end{equation}
where $n^\mu$ is the outward-pointing normal vector to $\partial M$ with unit norm and $h$ is the determinant of the induced metric. In our case $n_\mu = -z^{-1} \delta^z_\mu$, hence
\begin{equation}
  \delta\mathcal{S}_{\mathrm{bulk}} = \frac{1}{16\pi G^{(5)}_\mathrm{N}} \int_{\partial M} d^4x \, \sqrt{-h} \, \bigl( 6X^{-2} z \, \partial_z X \, \delta X \bigr) \,.
\end{equation}
Finally, we have
\begin{equation}
	\frac{1}{\sqrt{-h}} \frac{\delta \mathcal{S}_\mathrm{reg}}{\delta X} = \frac{1}{16\pi G^{(5)}_\mathrm{N}} \bigl( 6X^{-2} \epsilon \, \partial_\epsilon X \bigr) \,.
\end{equation}
The counterterm \eqref{S-count} gives
\begin{equation}
	\frac{1}{\sqrt{-h}} \frac{\delta \mathcal{S}_\mathrm{c.t.}}{\delta X} = \frac{1}{16\pi G^{(5)}_\mathrm{N}} \biggl[ 12(1-X) + \frac{6}{\log\epsilon} (1-X) \biggr] \,.
\end{equation}
We can finally compute the one-point function for the scalar, obtaining the following result:
\begin{equation} \label{vev_X}
	\langle\mathcal{O}_X\rangle = \frac{1}{16\pi G^{(5)}_\mathrm{N}} \lim_{\epsilon\to0} \biggl[ \frac{\log\epsilon}{\epsilon^2} \biggl( -\frac{\epsilon^2}{\log\epsilon} \, 6X_2 + \mathcal{O}(\epsilon^3) \biggr) \biggr] = \frac{1}{16\pi G^{(5)}_\mathrm{N}} (-6X_2) \,.
\end{equation}
We can thus observe that the VEV of the operator dual to the scalar field corresponds to the $z^2$ term in the FG expansion of~$X$, as expected from its conformal dimension. We recall that, in general, the one-point functions depend on all the sources and that the VEVs are obtained setting the sources to zero.

The one-point function of the current dual to the two-form is
\begin{equation}
	\langle J^{ij}_\alpha \rangle = \frac{1}{\sqrt{-g_0}} \frac{\delta \mathcal{S}_\mathrm{ren}}{\delta B^\alpha_{-1\,ij}} =
	\lim_{\epsilon\to0} \biggl( \frac{1}{\epsilon^5} \frac{1}{\sqrt{-h}} \frac{\delta \mathcal{S}_\mathrm{sub}}{\delta B^\alpha_{ij}} \biggr) \,,
\end{equation}
while the one-point function of the stress--energy tensor of the dual theory is defined as
\begin{equation}
	\langle T_{ij} \rangle = \frac{-2}{\sqrt{-g_0}} \frac{\delta \mathcal{S}_\mathrm{ren}}{\delta g_0^{ij}} = \lim_{\epsilon\to0} \biggl( \frac{1}{\epsilon^2} \frac{-2}{\sqrt{-h}} \frac{\delta \mathcal{S}_\mathrm{sub}}{\delta h^{ij}} \biggr) = \lim_{\epsilon\to0} \biggl( \frac{1}{\epsilon^2} \, T_{ij}[h] \biggr) \,,
\end{equation}
where $T_{ij}[h]$ is the boundary stress--energy tensor. This tensor can be decomposed as
\begin{equation}
	T_{ij}[h] = T^\mathrm{reg}_{ij}[h] + T_{ij}^\mathrm{c.t.}[h] \,,  \qquad  \text{with}  \qquad
	T^\mathrm{reg}_{ij}[h] = -\frac{1}{8\pi G^{(5)}_\mathrm{N}} \bigl( K_{ij} - K h_{ij} \bigr) \,,
\end{equation}
where $K=h^{kl} K_{kl}$ is the trace of the extrinsic curvature tensor. Unfortunately, these last two correlators cannot be derived explicitly from the renormalized action that we presented. This is due to the fact that, to construct the counterterms, we made use of the equations of motion of the theory, that for our solution give rise to a constraint (see equation \eqref{BB=0}) among the matter fields. Even if this condition greatly simplifies the computation, it also causes the loss of some information, making impossible to recover the off-shell renormalized action, needed to compute the dual stress--energy tensor.
We expect that this issue would be overcome if we considered the more general expression for the action, without truncating the vectors $A^i, \mathcal{A}$ to zero. Such a computation is beyond the scope of the present work.

\section{Conclusions}\label{conclusions}

In this paper we constructed new families of $\AdS_2 \times S^2 \times I$ solutions to the five-dimensional $\mathcal{N}=4$ $\SU{2}\times\U{1}$ Romans' supergravity, characterized by non-vanishing two-forms and a real scalar field . These backgrounds are supersymmetric, in particular they preserve half of the total number of supercharges, and exhibit different behaviors according to the scaling of the $\AdS_2$ and the $S^2$ warp factors. When they have the same radius, the solution describes a domain wall interpolating between an asymptotically locally $\AdS_5$ geometry and a spacetime singularity. When the two spaces scale differently, the resulting geometry is a regular Janus that interpolates between $\AdS_5$ and a smooth $\AdS_2 \times \RR^3$ geometry. Remarkably, the latter system can be written as a one-parameter family of deformations of the global $\AdS_5$, which is recovered when the parameter is set to zero.

By means of the truncation Ansatz developed in~\cite{Lu:1999bw}, we can uplift our solutions to $D=10$ and embed them in Type~IIB supergravity. Both classes give rise to BPS/2 configurations described by $\AdS_2 \times S^2 \times S^1 \times S^3$ geometries fibered over two intervals, with $\AdS_5 \times S^5$ asymptotics. Additionally, the three-sphere can be replaced by its orbifold counterpart $S^3/\ZZ_k$ as in~\eqref{orbifoldS5}, further halving the number of preserved supercharges and yielding an $\AdS_5 \times S^5/\ZZ_k$ asymptotic geometry.
The 10D solutions present the same behavior of the lower-dimensional geometries: while the 5D configurations with equal radii give rise to singular domain walls in Type IIB, the uplifted Janus solutions are smooth everywhere.

The uplifted solution allows us to provide the Type~IIB origin of the 5D domain wall background. In particular, the resulting geometry can be engineered intersecting a stack of D3 branes, with the addition of KK monopoles, with a suitable bound state of defect D1-F1-D5-NS5 branes, as first presented and discussed in~\cite{Lozano:2021fkk}. In fact, through an appropriate change of coordinates, the near-horizon of the latter~\eqref{nearHorizonTypeIIB} can be shown to perfectly match with the uplifted domain wall~\eqref{chargedDWUplift}.
A key aspect in this construction is that the charges of D1-F1-D5-NS5 branes are smeared along the transverse directions of D3 branes, which, as final result, implies that the D3 branes worldvolume is deformed by the defect branes to the product $\AdS_2 \times S^2$, the two of them having equal radii.
On the other hand, we were not able to provide the brane picture underlying our Janus solution. However, the similarities shared with the domain wall seems to suggest that the brane setup could be the same, but with the defect branes fully-localized in the D3 branes transverse space.

In the last section of the paper we study our Janus solutions holographically. Specifically, we compute the on-shell action and the one-point function. To do so, we apply the standard techniques of holographic renormalization.

This work opens up several potential directions for future exploration. Below, we outline three of them. The first immediate research trajectory is to study in detail the superconformal quantum mechanics associated with the line defect of the solutions \eqref{5DJanus}. This direction has already been explored for similar AdS$_2$ backgrounds in string theory with the same amount of supersymmetry \cite{Lozano:2020sae,Lozano:2020txg,Lozano:2021fkk,Lozano:2022vsv,Conti:2023naw}. To this aim, a very important step forward would be providing a clear brane picture of the uplift \eqref{JanusUplift} of our 5D Janus. In relation to this last point, a second interesting direction involves investigating possible connections between line defects and black holes. Specifically, our $\AdS_2 \times S^2 \times I$ regular solution should be embedded within a more general 5D solution, where the bulk excitations also depend on the radial coordinate of AdS$_2$. Such a solution might represent a novel type of AdS$_5$ black hole.

Finally, a third intriguing direction is to explore potential connections with recent developments in the Swampland program. In \cite{Li:2023gtt}, a notion of distance between stringy AdS solutions is introduced. It would be interesting to test this framework with variations over AdS$_5 \times S^5$ vacua expressed as deformations of the geometry $\AdS_2 \times S^2$, and with the contribution of flux variations associated to the two-form fields.

\section*{Acknowledgements}

We would like to thank Antonio Amariti, Giuseppe Dibitetto, Niall Macpherson for interesting discussions and comments.
The work of NP is supported by INFN through the project ``Understanding gravity via gauge theories, supergravity and strings''.

\appendix

\section{Spinor conventions}
\label{app:gamma-matr}

In this appendix we introduce the representations used for gamma matrices. Moreover, we summarize the main features of symplectic-Majorana spinors in five dimensions. As a general rule, we use Greek letters for curved indices and Latin indices for the flat bases. Given the general prescription \eqref{generalMetricAnsatz} for the 5D geometry, the curved coordinates are given by $x^\mu=(t,u,\rho,\theta_1,\theta_2)$. The AdS$_2$ and $S^2$ metrics are then taken as it follows
\begin{equation}
	d s_{\AdS_2}^2 = -u^2 \, d t^2 + \frac{d u^2}{u^2}  \qquad \text{and} \qquad
	d s_{S^2}^2 = d\theta_1^2 + \sin^2\!\theta_1 \, d\theta_2^2 \,.
\end{equation}
The flat basis is then labeled by Latin indices $a=0,\ldots,5$. We also introduce 2D bases for AdS$_2$ and $S^2$. For corresponding curved and flat indices we use the subscript $\hat \mu=(t,u)$ and $\hat a=(0,1)$ for AdS$_2$ indices, and $\tilde \mu=(\theta_1, \theta_2)$ and $\tilde a=(1,2)$ for $S^2$.

\subsection{Gamma matrix representations}

The 5D Lorentzian gamma matrices $\gamma_a$ satisfy the fundamental relation $\{\gamma_a,\gamma_b\} = 2\eta_{ab}$. For these matrices we choose the following decomposition~\cite{Lu:1998nu}
\begin{equation} \label{gamma-deco}
	\gamma_{\hat a} = \beta_{\hat a} \otimes \rho_* \,,  \qquad\quad  \gamma_2 = \beta_* \otimes \rho_* \,,  \qquad\quad  \gamma_{\tilde a} = \mathbb{I}_2 \otimes \rho_{\tilde a} \,,
\end{equation}
where $\beta_{\hat{a}}$ are the 2D Lorentzian gamma matrices and $\rho_{\tilde{a}}$ the 2D Euclidean gamma matrices. The chiral matrices are defined as $\beta_* = -\beta_0 \beta_1$ and $\rho_* = -i\rho_1\rho_2$.
For lower-dimensional gamma matrices, $\beta_{\hat a}$ and $\rho_{\tilde a}$, we use the following representation in terms of the Pauli matrices $\sigma^{1,2,3}$,
\begin{equation} \label{gamma-rep}
	\beta_0 = i \sigma^1 \,,  \qquad  \beta_1 = \sigma^2 \,,  \qquad \text{and}\qquad
	\rho_1 = \sigma^1 \,,  \qquad  \rho_2 = \sigma^2 \,,
\end{equation}
which gives $\beta_* = \sigma^3$ and $\rho_* = \sigma^3$ . This choice implies that $\gamma_{01234} = -i \,\mathbb{I}_4$.

In SUSY variations \eqref{susy-var} gamma matrices $\Gamma_I$ associated to the $R$-symmetry group $\Spin{5}_R \simeq \USp{4}_R$ appear. These are five-dimensional Euclidean gamma matrices satisfying the relation $\{\Gamma_I,\Gamma_J\} = 2\delta_{IJ}$. In the SUSY variations the index $I$ is then split as $I=(i,\alpha)$.
A useful representation for these matrices is
\begin{equation} \label{Gamma-rep}
	\begin{gathered}
		\Gamma_1 = -\sigma^1 \otimes \mathbb I_2 \,,  \qquad  \Gamma_2 = -\sigma^2 \otimes \mathbb I_2 \,,  \qquad
		\Gamma_3 = -\sigma^3 \otimes \sigma^1 \,, \\
		\Gamma_4 = -\sigma^3 \otimes \sigma^2 \,,  \qquad  \Gamma_5 = -\sigma^3 \otimes \sigma^3 \,.
	\end{gathered}
\end{equation}
With this choice the skew-symmetric matrix $\Omega_{xy}$ used to raise the $\USp{4}_R$ index~$x$ (see~\cite{Romans:1985ps} for more details) takes the explicit form
\begin{equation}
	\Omega^{xy} = \Omega_{xy} = i \sigma^1 \otimes \sigma^2 \,.
\end{equation}
For the above choice, one has that $\Gamma_{12345} = \mathbb{I}_4$.

\subsection{Symplectic-Majorana spinors in $D=5$}

The components of the SUSY parameter $\epsilon_x$ in \eqref{susy-var} are organized in $\ma N=4$ symplectic-Majorana spinors. These are defined by the following condition
\begin{equation} \label{Maj-cond}
	\epsilon_x^* = \Omega^{xy} \mathcal{B}\, \epsilon_y \,,
\end{equation}
where $x, y$ are $\USp{4}_R$ indices and the matrix $\Omega^{xy}$ was introduced above.
The matrix $\mathcal{B}$ is related to the five-dimensional charge conjugation matrix $\mathcal{C}$ by the condition $\mathcal{B}=i\,\mathcal{C}\gamma^0$.
Given the decomposition \eqref{gamma-deco}, consistency requires matrix $\mathcal{B}$ to splits as\footnote{If we choose to have Majorana spinors also along $S^2$, then we have $\mathcal{B} = \mathcal{B}_2^\beta \otimes (\mathcal{B}_2^\rho \rho_*)$. However, this possibility does not seem to be consistent.}
\begin{equation}\label{Bdecomposition}
	\mathcal{B} = \mathcal{B}_2^\beta \otimes \mathcal{B}_2^\rho \,.
\end{equation}
We may recall the properties that the $\mathcal{B}$ matrix has to satisfy~\cite{VanProeyen:1999ni}
\begin{itemize}
\item $D=5$ (Lorentzian)
\begin{equation}
	(\gamma_a)^* = -\mathcal{B} \gamma_a \mathcal{B}^{-1} \,,  \qquad  \mathcal{B}^* \mathcal{B} = -\mathbb I_4 \,.
\end{equation}

\item $D=2$ (Lorentzian)
\begin{equation}
	\begin{aligned}
		(\beta_{\hat a})^* &= \mathcal{B}_2^\beta \beta_{\hat a} (\mathcal{B}_2^{\beta})^{-1} \,,  \qquad  & (\mathcal{B}_2^{\beta})^* \mathcal{B}_2^\beta &= +\mathbb I_2 \,, \\
		\beta_* &= -\beta_0 \beta_1 \,,  \qquad  & (\mathcal{B}_2^\beta \beta_*)^* &= (\mathcal{B}_2^\beta \beta_*)^{-1} \,.
	\end{aligned}
\end{equation}

\item $D=2$ (Euclidean)
\begin{equation}
	\begin{aligned}
		(\rho_{\tilde a})^* &= -\mathcal{B}_2^\rho \,\rho_{\tilde a}\,(\mathcal{B}_2^\rho)^{-1} \,,  \qquad  & (\mathcal{B}_2^\rho)^*\, \mathcal{B}^\rho &= -\mathbb I_2 \,, \\
		\rho_* &= -i \rho_1 \rho_2 \,,  \qquad  & (\mathcal{B}^\rho_2 \,\rho_*)^* &= (\mathcal{B}_2^\rho\, \rho_*)^{-1} \,.
	\end{aligned}
\end{equation}
\end{itemize}

\section{The construction of counterterms}
\label{app:counterterms}

In this appendix, we summarize the computation of the counterterms required by the holographic renormalization procedure.
We begin assuming a Fefferman--Graham expansion of the fields, as in equations~\eqref{FG-deco}--\eqref{XB-deco}.
As a consequence, the five-dimensional Ricci tensor decomposes as~\cite{Henningson:1998gx,deHaro:2000vlm}
\begin{align}
	R_{ij} &= R_{ij}[g] + \frac12 g'_{ik} g^{kl} g'_{lj} - \frac14 g'_{ij} \tr(g^{-1} g') - \frac12 g_{ij}'' + \frac12 \, z^{-1} \bigl( 3g'_{ij} + g_{ij} \tr(g^{-1} g') \bigr) - 4z^{-2} g_{ij} \,, \nonumber \\
	R_{iz} &= \frac12 g^{jk} \nabla_k g'_{ij} - \frac12 g^{jk} \nabla_i g'_{jk} \,, \\[0.2em]
	R_{zz} &= \frac14 \tr(g^{-1} g' g^{-1} g') - \frac12 \tr(g^{-1} g'') + \frac12 \, z^{-1} \tr(g^{-1} g') - 4z^{-2} \nonumber \,,
\end{align}
where $R_{ij}[g]$ and $\nabla_i$ are constructed using the lower-dimensional metric $g$ and the prime denotes the derivative with respect to~$z$.
In the derivation of counterterms, we will use extensively the formula
\begin{equation}
	\begin{split}
		\sqrt{-g} &= \sqrt{-g_0} \biggl[ 1 + \frac12 z^2 \tr(g_0^{-1} g_2) + \frac12 z^4 \biggl( -\frac12 \tr(g_0^{-1} g_2 g_0^{-1} g_2) + \frac14 \tr^2(g_0^{-1} g_2) \\
		& + \tr(g_0^{-1} g_4) + \log z \tr(g_0^{-1} \tilde{g}_4) + \log^2\!z \tr(g_0^{-1} \hat{g}_4) \biggr) \biggr] + \mathcal{O}(z^5) \,,
	\end{split}
\end{equation}
representing the expansion of the determinant of the metric.

\subsection{The order-by-order equations of motion}

A key feature of the Fefferman--Graham expansion is that most of the higher-order contributions are fixed by leading-order ones through the equations of motion. These relations are obtained expanding the equations of motion in series of~$z$ and imposing them to vanish at each order.

Since we are not interested in constructing the most general set of counterterms, but rather only the ones that fit into our Ansatz, in what follows, we will truncate to zero the vector degrees of freedom of the Romans supergravity, namely $A^i = \mathcal{A} = 0$. This choice is consistent provided that the second Maxwell equation~\eqref{maxwell-eq} is satisfied. This is equivalent to impose the constraint
\begin{equation}
	B \wedge \bar{B} = \sum_\alpha B^\alpha \wedge B^\alpha = 0 \,.
\end{equation}
Assuming this truncation, from the $(zz)$-component of the Einstein equations, expanded up to order $\mathcal{O}(z^2)$, we obtain
\begin{equation} \label{BB=0}
	\sum_\alpha |B^\alpha_{-1}|^2_{g_0} \equiv \sum_\alpha \frac{1}{2} \, g_0^{ik} g_0^{jl} B^\alpha_{-1\,ij} B^\alpha_{-1\,kl} = 0 \,,
\end{equation}
and the following series of trace relations
\begin{equation} \label{trg04}
	\begin{split}
		\tr(g_0^{-1} g_4) &= \frac14 \tr(g_0^{-1} g_2 g_0^{-1} g_2) - 2X_2^2 - \frac14 \tilde{X}_2^2 - \frac{1}{24} \tr(g_0^{-1} B^\alpha_{-1} g_0^{-1} B^\alpha_1) \\
		& + \frac{1}{32} \tr(g_0^{-1} B^\alpha_{-1} g_0^{-1} \tilde{B}^\alpha_1) + \frac{1}{24} \tr(g_0^{-1} B^\alpha_{-1} g_0^{-1} B^\alpha_{-1} g_0^{-1} g_2) \,, \\
		\tr(g_0^{-1} \tilde{g}_4) &= -4X_2 \tilde{X}_2 - \frac{1}{24} \tr(g_0^{-1} B^\alpha_{-1} g_0^{-1} \tilde{B}^\alpha_1) \,, \\
		\tr(g_0^{-1} \hat{g}_4) &= -2\tilde{X}_2^2 \,.
	\end{split}
\end{equation}
The $(ij)$-components of the Einstein equations, together with~\eqref{BB=0}, gives (sum over $\alpha$ is understood)
\begin{gather}
	\label{g2}
	g_{2\,ij} = -\frac12 \biggl( R_{ij}[g_0] - \frac16 \, g_{0\,ij} R[g_0] \biggr) - \frac14 B^\alpha_{-1\,ik} g_0^{kl} B^\alpha_{-1\,lj} \,, \\
	\label{trg02}
	\tr(g_0^{-1} g_2) = -\frac16 R[g_0] \,.
\end{gather}
The equation for the scalar does not provide any additional constraint, whereas Maxwell's equations impose, among the others, the relations
\begin{equation}
	dB^\alpha_{-1} = dB^\alpha_1 = d\tilde{B}^\alpha_1 = 0 \,.
\end{equation}
One last constraint involving the components of the two-form can be obtained from the ``square'' of the last Maxwell equation in~\eqref{maxwell-eq}, namely
\begin{equation} \label{os-max}
	\star d\bigl( X^2 \star\! dB^\alpha \bigr) = - X^{-2} B^\alpha \,.
\end{equation}
Indeed, plugging the FG expansion of $B^\alpha$, equation~\eqref{os-max} implies
\begin{equation} \label{B1t}
	\tilde{B}^\alpha_{1\,ij} = \frac12 \tr(g_0^{-1} g_2) B^\alpha_{-1\,ij} + \tilde{X}_2 B^\alpha_{-1\,ij} - g_{2\,ik} g_0^{kl} B^\alpha_{-1\,lj} - B^\alpha_{-1\,ik} g_0^{kl} g_{2\,lj} \,.
\end{equation}
We close this section with a convenient expression for the Ricci scalar $R[g]$, obtained contracting the $(ij)$-components of the Einstein equations with $g^{ij}$:
\begin{equation} \label{Rg}
  R[g] = R[g_0] + z^2 \Bigl[ 2\tr(g_0^{-1} g_2 g_0^{-1} g_2) + \tr^2(g_0^{-1} g_2) - \frac14 \tr(g_0^{-1} B^\alpha_{-1} g_0^{-1} \tilde{B}^\alpha_1) \Bigr] + \mathcal{O}(z^3) \,.
\end{equation}
In writing this relation, we already made use of the constraints presented above.

\subsection{The counterterms}

Now that we have at our disposal the essential relations involving the coefficients in the FG expansion, we are ready for the construction of the counterterms. As we explained in section~\ref{sec:holo-ren}, the strategy is to set a cutoff at $z=\epsilon$ and to compute the regularized action, obtained integrating the bulk action up to the cutoff and evaluating the GHY term at the cutoff. The action will then present power-law and logarithmic divergences in $\epsilon$. The latter due to the fact that the boundary is even-dimensional.
It is important to notice that a given counterterm could introduce new, less severe divergences, and therefore will have to be taken into account in the rest of the analysis.

The starting point is the bulk action~\eqref{action}, truncated to $A^i = \mathcal{A} = 0$ and with $g_1=1$, $g_2=\sqrt2$. By means of the equations of motion, the on-shell action can be written in the following form
\begin{equation}
	\mathcal{S}_\mathrm{bulk} = \frac{1}{16\pi G^{(5)}_\mathrm{N}} \int_M d^5x \, \sqrt{-G} \, \biggl( \frac23 \, \mathcal{V} + \frac{1}{12} X^{-2} B^\alpha_{\mu\nu} B^{\alpha\,\mu\nu} \biggr) \,.
\end{equation}
The GHY term~\eqref{GHYterm} comprises the induced boundary metric~$h_{ij}$ and its extrinsic curvature~$K_{ij}$. In the FG coordinates they read as
\begin{equation}
	h_{ij} = z^{-2} g_{ij} \,,  \qquad  K_{ij} = -\frac{z}{2} \, \partial_z h_{ij} \,,
\end{equation}
thus yielding a GHY term of the form
\begin{equation}
	\mathcal{S}_\mathrm{GHY} = \frac{1}{16\pi G^{(5)}_\mathrm{N}} \int_{\partial M} d^4x \, \bigl( -2z \, \partial_z \sqrt{-h} \bigr) \,.
\end{equation}
Let's write the various divergent contributions produced in our specific case. To keep the notation light, in what follows we will write the various terms in a schematic form, omitting the prefactor $\frac{1}{16\pi G^{(5)}_\mathrm{N}}\int d^4x\sqrt{-g_0}$ and writing just the non-trivial coefficient.
\paragraph{Order $\boldsymbol{\epsilon^{-4}}$:}
The most severe divergences that we encounter appear at order $\mathcal{O}(\epsilon^{-4})$ and are given by
\begin{align*}
	& \mathcal{S}_\mathrm{pot}: && -2 \\
	& \mathcal{S}_\mathrm{GHY}: && 8
\end{align*}
The total divergence is then
\begin{equation}
	\mathcal{S}_\mathrm{div}^{(4)} = \frac{\epsilon^{-4}}{16\pi G^{(5)}_\mathrm{N}}  \int d^4x \, \sqrt{-g_0} \, \bigl[ 6 \bigr] \,,
\end{equation}
which can be canceled by the covariant counterterm
\begin{equation}
	\mathcal{S}_\mathrm{c.t.}^{(4)} = \frac{1}{16\pi G^{(5)}_\mathrm{N}}  \int_{\partial M} d^4x \, \sqrt{-h} \, \bigl[ -6 \bigr] \,.
\end{equation}

\paragraph{Order $\boldsymbol{\epsilon^{-2}}$:}
Due to our specific expansion, we do not have divergences at order $\mathcal{O}(\epsilon^{-3})$, therefore the following ones are at $\mathcal{O}(\epsilon^{-2})$
\begin{align*}
	& \mathcal{S}_\mathrm{pot}: && -2\tr(g_0^{-1} g_2) \\
	& \mathcal{S}_{BB}: && \frac{1}{12} |B^\alpha_{-1}|_{g_0}^2 \\
	& \mathcal{S}_\mathrm{GHY}: && 2\tr(g_0^{-1} g_2) \\
	& \mathcal{S}_\mathrm{c.t.}^{(4)}: && -3\tr(g_0^{-1} g_2)
\end{align*}
Making use of~\eqref{BB=0} and~\eqref{trg02}, the total divergence can be written as
\begin{equation}
	\mathcal{S}_\mathrm{div}^{(2)} = \frac{\epsilon^{-2}}{16\pi G^{(5)}_\mathrm{N}}  \int d^4x \, \sqrt{-g_0} \, \biggl[ \frac12 R[g_0] \biggr] \,,
\end{equation}
and the corresponding covariant counterterm is
\begin{equation}
	\mathcal{S}_\mathrm{c.t.}^{(2)} = \frac{1}{16\pi G^{(5)}_\mathrm{N}}  \int_{\partial M} d^4x \, \sqrt{-h} \, \biggl[ -\frac12 R[h] \biggr] \,.
\end{equation}
In order to compute the subsequent divergences introduced by this counterterm, relation~\eqref{Rg} will be essential.

\paragraph{Order $\boldsymbol{\log^2\!\epsilon}$:}
Since order $\mathcal{O}(\epsilon^{-1})$ gives no divergences, we move to the family of logarithmic ones. The first examples appear at order $\mathcal{O}(\log^3\!\epsilon)$, but they cancel when~\eqref{trg04} is imposed. The next divergences, at order $\mathcal{O}(\log^2\!\epsilon)$, read
\begin{align*}
	& \mathcal{S}_\mathrm{pot}: && 2\tr(g_0^{-1} \tilde{g}_4) + 8X_2 \tilde{X}_2 \\
	& \mathcal{S}_{BB}: && \frac{1}{12} \tr(g_0^{-1} B^\alpha_{-1} g_0^{-1} \tilde{B}^\alpha_1) + \frac16 \tilde{X}_2 |B^\alpha_{-1}|_{g_0}^2 \\
	& \mathcal{S}_\mathrm{c.t.}^{(4)}: && -3\tr(g_0^{-1} \hat{g}_4)
\end{align*}
Using again~\eqref{BB=0} and~\eqref{trg04}, the total contribution becomes
\begin{equation}
	\mathcal{S}_\mathrm{div}^{(\ell.2)} = \frac{\log^2\!\epsilon}{16\pi G^{(5)}_\mathrm{N}}  \int d^4x \, \sqrt{-g_0} \, \bigl[ 6\tilde{X}_2^2 \bigr] \,,
\end{equation}
which can be canceled by the counterterm
\begin{equation}
	\mathcal{S}_\mathrm{c.t.}^{(\ell.2)} = \frac{1}{16\pi G^{(5)}_\mathrm{N}}  \int_{\partial M} d^4x \, \sqrt{-h} \, \bigl[ -6(1 - X)^2 \bigr] \,.
\end{equation}

\paragraph{Order $\boldsymbol{\log\epsilon}$:}
The last set of divergences appear at order $\mathcal{O}(\log\epsilon)$ and have the form
\begin{align*}
	& \mathcal{S}_\mathrm{pot}: && \tr^2(g_0^{-1} g_2) - 2\tr(g_0^{-1} g_2 g_0^{-1} g_2) + 4\tr(g_0^{-1} g_4) + 8X_2^2 \\
	& \mathcal{S}_{BB}: && \frac16 \tr(g_0^{-1} B^\alpha_{-1} g_0^{-1} B^\alpha_1) - \frac16 \tr(g_0^{-1} B^\alpha_{-1} g_0^{-1} B^\alpha_{-1} g_0^{-1} g_2) + \frac13 X_2 |B^\alpha_{-1}|_{g_0}^2 - \frac{1}{12} \tr(g_0^{-1} g_2) |B^\alpha_{-1}|_{g_0}^2 \\
	& \mathcal{S}_\mathrm{GHY}: && -2\tr(g_0^{-1} \hat{g}_4) \\
	& \mathcal{S}_\mathrm{c.t.}^{(4)}: && -3\tr(g_0^{-1} \tilde{g}_4) \\
	& \mathcal{S}_\mathrm{c.t.}^{(\ell.2)}: && -12X_2 \tilde{X}_2
\end{align*}
All these contributions sum up to
\begin{equation}
	\mathcal{S}_\mathrm{div}^{(\ell.1)} = \frac{\log\epsilon}{16\pi G^{(5)}_\mathrm{N}}  \int d^4x \, \sqrt{-g_0} \, \biggl[ \tr^2(g_0^{-1} g_2) - \tr(g_0^{-1} g_2 g_0^{-1} g_2) + 3\tilde{X}_2^2 + \frac14 \tr(g_0^{-1} B^\alpha_{-1} g_0^{-1} \tilde{B}^\alpha_1) \biggr] \,.
\end{equation}
Using the explicit form of $g_2$ and $\tilde{B}^\alpha_1$ in~\eqref{g2} and~\eqref{B1t}, it is possible to show that this logarithmic divergence can be reabsorbed by the following counterterm
\begin{equation}
  \begin{split}
    \mathcal{S}_\mathrm{c.t.}^{(\ell.1)} & = \frac{1}{16\pi G^{(5)}_\mathrm{N}} \int_{\partial M} d^4x \, \sqrt{-h} \, \biggl\{ -\log\epsilon \biggl[ \frac{1}{12} R[h]^2 - \frac14 R_{ij}[h] R^{ij}[h] \\
    & + \frac{1}{16} \tr\bigl[(h^{-1} B^\alpha)^2 (h^{-1} B^\beta)^2\bigr] \biggr] - \frac{3}{\log\epsilon}(1-X)^2 \biggr\} \, .
  \end{split}
\end{equation}
Summing all the counterterms corresponding to the different orders of divergence, we obtain
\begin{equation}
  \begin{split}
    \ma S_{\mathrm{c.t.}} & = \frac{1}{16\pi G^{(5)}_\mathrm{N}}  \int_{\partial M} d^4x \, \sqrt{-h} \, \biggl\{ -6 - \frac12 R[h] -6(1-X)^2 - \log\epsilon \biggl[ \frac{1}{12} R[h]^2 \\
    & - \frac14 R_{ij}[h] R^{ij}[h] + \frac{1}{16} \tr\bigl[(h^{-1} B^\alpha)^2 (h^{-1} B^\beta)^2\bigr] \biggr] - \frac{3}{\log\epsilon}(1-X)^2 \biggr\} \, ,
  \end{split}
\end{equation}
where all the quantities are to be evaluated at the cutoff.


 \bibliographystyle{utphys}
  \bibliography{references}
\end{document}